\documentclass[12pt]{article}
\pdfoutput=1
\usepackage{jheppub}
\usepackage{epsfig}
\usepackage{amsmath}
\usepackage{amssymb}
\usepackage{amsfonts}
\usepackage{amsxtra}
\usepackage{amsthm}
\usepackage{mathrsfs}
\usepackage{makeidx}
\usepackage{graphics}
\usepackage{dsfont}
\usepackage{mathtools}
\usepackage{graphicx}
\usepackage{subcaption}
\usepackage{placeins}
\usepackage{bm}
\usepackage[capitalise]{cleveref}
\usepackage{empheq}
\usepackage{colortbl}
\usepackage{xcolor}
\usepackage{enumerate}
\usepackage{titlesec}
\usepackage{longtable}
\usepackage{float}
\usepackage{color}
\usepackage{tikz}
\usepackage{xfrac}
\usepackage{footnote}
\usepackage{rotating}
\usepackage{lscape}
\usepackage{makecell}
\usepackage{environ}
\usepackage{tabularx}
\usepackage{subfiles}
\usepackage[export]{adjustbox}
\usepackage{ytableau}
\usepackage{tikz-3dplot}
\usepackage{slashed}
\usepackage{pifont}
\usepackage{multirow}
\usepackage{mdframed}
\usepackage{bbm}
\usepackage[normalem]{ulem}
\usepackage{enumitem}

\usetikzlibrary{positioning,trees,decorations.pathmorphing,decorations.markings,decorations.pathreplacing,calc,shapes,patterns,arrows,chains,arrows.meta,fit,fadings,decorations.markings,graphs,graphs.standard,quotes}

\titleformat{\paragraph}
{\normalfont\normalsize\bfseries}{\theparagraph}{1em}{}
\titlespacing*{\paragraph}{0pt}{3.25ex plus 1ex minus .2ex}{1.5ex plus .2ex}

\DeclareMathOperator{\U}{U}
\DeclareMathOperator{\SU}{SU}
\DeclareMathOperator{\SO}{SO}

\DeclareMathOperator{\USp}{USp}

\DeclareMathOperator{\rank}{rank}

\newcommand{\coma}{\, , \quad}
\newcommand{\fstop}{\, .}

\def\ZZ{{\mathbb{Z}}}

\def\IX{{\bf {X}}}

\theoremstyle{definition}

\catcode`\@=11   
\newdimen\@rotdimen
\newbox\@rotbox  

\def\@vspec#1{\special{ps:#1}}
\def\@rotstart#1{\@vspec{gsave currentpoint currentpoint translate
   #1 neg exch neg exch translate}}
\def\@rotfinish{\@vspec{currentpoint grestore moveto}}
\def\@rotr#1{\@rotdimen=\ht#1\advance\@rotdimen by\dp#1%
   \hbox to\@rotdimen{\hskip\ht#1\vbox to\wd#1{\@rotstart{90 rotate}%
   \box#1\vss}\hss}\@rotfinish}
\def\@rotl#1{\@rotdimen=\ht#1\advance\@rotdimen by\dp#1%
   \hbox to\@rotdimen{\vbox to\wd#1{\vskip\wd#1\@rotstart{270 rotate}%
   \box#1\vss}\hss}\@rotfinish}%
\def\@rotu#1{\@rotdimen=\ht#1\advance\@rotdimen by\dp#1%
   \hbox to\wd#1{\hskip\wd#1\vbox to\@rotdimen{\vskip\@rotdimen
   \@rotstart{-1 dup scale}\box#1\vss}\hss}\@rotfinish}%
\def\@rotf#1{\hbox to\wd#1{\hskip\wd#1\@rotstart{-1 1 scale}%
   \box#1\hss}\@rotfinish}%
\def\rotate{\@ifnextchar[{\@rotate}{\@rotate[l\right]}}
\def\@rotate[#1]#2{\setbox\@rotbox=\hbox{#2}\@nameuse{@rot#1}\@rotbox}

\catcode`\@=12

\usetikzlibrary{positioning}
\usetikzlibrary{chains}
\usetikzlibrary{arrows, arrows.meta ,fit,decorations.pathreplacing}
\tikzstyle{every picture}+=[remember picture]
\tikzstyle{na} = [baseline]
\tikzstyle{ligne}=[draw, thick]

\usetikzlibrary{arrows, decorations.markings, calc, fadings, decorations.pathreplacing, patterns, decorations.pathmorphing, positioning}
\tikzset{>={Latex[width=1.5mm,length=1.5mm]}}
\tikzset{bd/.style={circle, draw=black, inner sep=0pt, fill=black, minimum size=1.2mm}}
\tikzset{bld/.style={circle, draw=blue, inner sep=0pt, fill=blue, minimum size=1.2mm}}
\tikzset{wd/.style={circle, draw=black, inner sep=0pt, fill=white, minimum size=1.2mm}}
\tikzset{rd/.style={circle, draw=red, inner sep=0pt, fill=red, minimum size=.9mm}}
\tikzset{wrd/.style={circle, draw=red, inner sep=0pt, fill=white, minimum size=.9mm}}
\usetikzlibrary{graphs,graphs.standard,quotes}

\def\node#1#2{\overset{#1}{\underset{#2}{{\color{gray} \bullet}}}}

\def\node#1#2{\overset{#1}{\underset{#2}{\circ}}}

\tikzstyle{every picture}+=[remember picture]
\tikzstyle{na} = [baseline=-.5ex]

\newcommand{\eg}{e.g. }

\newcommand{\ie}{i.e. }

\numberwithin{equation}{section}
\newcommand{\bes}[1]{\begin{equation} \begin{split} #1\end{split} \end{equation}}

\newcommand{\be}{\begin{equation}} \newcommand{\ee}{\end{equation}}
\newcommand{\bea}{\begin{equation} \begin{aligned}} \newcommand{\eea}{\end{aligned} \end{equation}}

\def\tilde{\widetilde}

\def\bar{\overline}

\def\rt2{\sqrt{2}}

\def\mod{{\rm mod}}

\def\CN{{\cal N}}

\def\CS{{\cal S}}

\def\1{{\ds 1}}

\newcommand{\fm}{\mathfrak{m}}
\newcommand{\fn}{\mathfrak{n}}

\def\SO{\mathrm{SO}}

\def\SU{\mathrm{SU}}

\def\Spin{\mathrm{Spin}}

\def\fN{\mathfrak{N}}

\def\repa{\raise4pt\hbox{$\square$}\mkern-14mu\raise-4pt\hbox{$\square$}}
\def\repab{\overline{\raise4pt\hbox{$\square$}\mkern-14mu\raise-4pt\hbox{$\square$}\mkern-1mu}}

\def\smileface{\ensuremath{\hbox{\large$\bigcirc$}\mkern-15mu\raise-1pt\hbox{\scriptsize$\smallsmile$}%
\mkern-10mu\raise4pt\hbox{..}\mkern4mu}}
\def\frownface{\ensuremath{\hbox{\large$\bigcirc$}\mkern-15mu\raise-1pt\hbox{\scriptsize$\smallfrown$}%
\mkern-10mu\raise4pt\hbox{..}\mkern4mu}}

\newcommand{\ba}{\begin{array}}
\newcommand{\ea}{\end{array}}
\newcommand{\bi}{\begin{itemize}}
\newcommand{\ei}{\end{itemize}}
\def\vec#1{\bm{#1}}
\def\bea#1\eea{\allowdisplaybreaks \begin{align}#1\end{align}}
 \newcommand{\ben}{\begin{enumerate}}
\newcommand{\een}{\end{enumerate}}
\newcommand{\bean}{\begin{eqnarray*}}
\newcommand{\eean}{\end{eqnarray*}}
\newcommand{\eref}[1]{(\ref{#1})}

\newcommand{\BZ}{\mathbb{Z}}

\definecolor{light-gray}{gray}{0.5}

\newcommand{\blue}{\color{blue}}

\newcommand{\red}{\color{red}}

\def\aup#1 {\overset{#1}{\uparrow} \, \overset{\tilde{#1}}{\downarrow}}

\tikzset{snake it/.style={decorate, decoration={snake, amplitude=.4mm, segment length=2mm,
                       post length=0mm,pre length=0mm}}}
                       
 \newcommand{\GCD}{\mathrm{GCD}}

\hypersetup{
	pdftitle={Dynamical consequences of 1-form symmetries and the exceptional Argyres-Douglas theories},    
	pdfauthor={\textcopyright\ Federico Carta, Simone Giacomelli, Noppadol Mekareeya, Alessandro Mininno},     
	pdfsubject={HEP},   
	pdfcreator={pdfLaTex},   
	pdfproducer={LaTex}, 
	pdfkeywords={},
	colorlinks=true,
}

\makeatletter
\newsavebox{\measure@tikzpicture}
\NewEnviron{scaletikzpicturetowidth}[1]{%
  \def\tikz@width{#1}%
  \begin{lrbox}{\measure@tikzpicture}%
  \BODY
  \end{lrbox}%
  \pgfmathparse{#1/\wd\measure@tikzpicture}%
  \BODY
}
\makeatother

\makeatletter
\def\squarecorner#1{
    \pgf@x=\the\wd\pgfnodeparttextbox%
    \pgfmathsetlength\pgf@xc{\pgfkeysvalueof{/pgf/inner xsep}}%
    \advance\pgf@x by 2\pgf@xc%
    \pgfmathsetlength\pgf@xb{\pgfkeysvalueof{/pgf/minimum width}}%
    \ifdim\pgf@x<\pgf@xb%
        \pgf@x=\pgf@xb%
    \fi%
    \pgf@y=\ht\pgfnodeparttextbox%
    \advance\pgf@y by\dp\pgfnodeparttextbox%
    \pgfmathsetlength\pgf@yc{\pgfkeysvalueof{/pgf/inner ysep}}%
    \advance\pgf@y by 2\pgf@yc%
    \pgfmathsetlength\pgf@yb{\pgfkeysvalueof{/pgf/minimum height}}%
    \ifdim\pgf@y<\pgf@yb%
        \pgf@y=\pgf@yb%
    \fi%
    \ifdim\pgf@x<\pgf@y%
        \pgf@x=\pgf@y%
    \else
        \pgf@y=\pgf@x%
    \fi
    \pgf@x=#1.5\pgf@x%
    \advance\pgf@x by.5\wd\pgfnodeparttextbox%
    \pgfmathsetlength\pgf@xa{\pgfkeysvalueof{/pgf/outer xsep}}%
    \advance\pgf@x by#1\pgf@xa%
    \pgf@y=#1.5\pgf@y%
    \advance\pgf@y by-.5\dp\pgfnodeparttextbox%
    \advance\pgf@y by.5\ht\pgfnodeparttextbox%
    \pgfmathsetlength\pgf@ya{\pgfkeysvalueof{/pgf/outer ysep}}%
    \advance\pgf@y by#1\pgf@ya%
}
\makeatother

\pgfdeclareshape{square}{
    \savedanchor\northeast{\squarecorner{}}
    \savedanchor\southwest{\squarecorner{-}}

    \foreach \x in {east,west} \foreach \y in {north,mid,base,south} {
        \inheritanchor[from=rectangle]{\y\space\x}
    }
    \foreach \x in {east,west,north,mid,base,south,center,text} {
        \inheritanchor[from=rectangle]{\x}
    }
    \inheritanchorborder[from=rectangle]
    \inheritbackgroundpath[from=rectangle]
}

\tikzset{stretch/.initial=1}
\newcommand\drawloop[4][]%
   {\draw[shorten <=0pt, shorten >=0pt,#1]
      ($(#2)!\pgfkeysvalueof{/tikz/stretch}!(#2.#3)$)
      let \p1=($(#2.center)!\pgfkeysvalueof{/tikz/stretch}!(#2.north)-(#2)$),
          \n1= {veclen(\x1,\y1)*sin(0.5*(#4-#3))/sin(0.5*(180-#4+#3))}
      in arc [start angle={#3-90}, end angle={#4+90}, radius=\n1]%
   }

\preprint{ZMP-HH/22-7}
\title{Dynamical consequences of 1-form symmetries and the exceptional Argyres-Douglas theories}
\author[a]{Federico Carta,}
\author[b,c]{~Simone Giacomelli,}
\author[b,c,d]{~Noppadol Mekareeya}
\author[e]{\\ and Alessandro Mininno}
\affiliation[a]{Department of Mathematical Sciences,
		Durham University, \\ Durham, DH1 3LE, United Kingdom}
\affiliation[b]{Dipartimento di Fisica, Universit\`a di Milano-Bicocca, Piazza della Scienza 3, \\ I-20126 Milano, Italy}
\affiliation[c]{INFN, sezione di Milano-Bicocca, Piazza della Scienza 3, \\ I-20126 Milano, Italy}
\affiliation[d]{Department of Physics, Faculty of Science, Chulalongkorn University, \\ Phayathai Road,
Pathumwan, Bangkok 10330, Thailand}
\affiliation[e]{II. Institut f\"ur Theoretische Physik, Universit\"at Hamburg,\\
Luruper Chaussee 149, 22607 Hamburg, Germany}
\emailAdd{federico.carta@durham.ac.uk}
\emailAdd{simone.giacomelli@unimib.it}
\emailAdd{n.mekareeya@gmail.com}
\emailAdd{alessandro.mininno@desy.de}
\abstract{Higher-form symmetries have proved useful in constraining the dynamics of a number of quantum field theories. In the context of the Argyres-Douglas (AD) theories of the $(G,G')$ type, we find that the 1-form symmetries are invariant under the Higgs branch flow, and that they are captured by the non-Higgsable sector at a generic point on the Higgs branch of the AD theory in question. As a consequence, dimensional reduction of an AD theory with a non-trivial 1-form symmetry to 3d leads to a free sector. We utilize these observations, along with other results, to propose systematically the mirror theories for the AD theories of the $(A_n, E_m)$ type. As a by-product of these findings, we discover many important results: the Flip-Flip duality for all $T[G]$ theories with simply-laced group $G$, including the exceptional ones; the class $\mathcal{S}$ descriptions of exceptional affine Dynkin diagram such that all gauge groups are special unitary; the universality of the mirror theories for $D_{h^\vee_G}(G)$ with $h^\vee_G$ the dual Coxeter number of $G$; and the triviality of the 2-group structure in the $(A_n, E_m)$ theories.}

\allowdisplaybreaks[1]

\begin{document} 

\maketitle

\section{Introduction}
\label{sec:intro}
Among the various quantum field theories, 4d $\CN=2$ superconformal field theories (SCFTs) play a distinguished role. The high amount of supersymmetry, together with the conformal invariance, allows for the possibility of obtaining exact results, and investigating strongly coupled regimes which are otherwise complicated to study.

One interesting class of $4$d $\mathcal{N}=2$ SCFTs consist of the so called Argyres-Douglas (AD) theories. They are characterized by the presence of Coulomb branch (CB) operators with fractional dimensions. AD theories were originally found to describe the low energy effective physics at special points in the CB of $4$d $\mathcal{N}=2$ gauge theories, such as for example pure $\SU(3)$ super-Yang-Mills \cite{Argyres:1995jj,Argyres:1995xn, Eguchi:1996vu, Eguchi:1996ds}. At such special loci, mutually non-local dyons become massless, and thus remain included in the low energy effective field theory. Due to this fact, Argyres-Douglas theories are intrinsically non-Lagrangian. Nevertheless, many AD theories can be realized and studied from a top-down point of view, often related to superstring theory. For example, a large class of AD theories can be realized inside class $\mathcal{S}$ \cite{Gaiotto:2009we, Gaiotto:2009hg, Xie:2012hs, Wang:2018gvb}, while another class of AD theories, with a partial overlap with the previous one, can be realized by geometric engineering of type IIB superstring theory \cite{Shapere:1999xr, Cecotti:2010fi}.

In this paper, we mainly focus on the geometric engineering setup. We consider Type IIB superstring theory compactified on a non-compact Calabi-Yau $3$-fold realized as the zero-locus of a single hypersurface equation in $\mathbb{C}^4$. In particular, we take the hypersurface equation to be of $(G,G')$ type \cite{Cecotti:2010fi}, namely given by the sum of two ADE polynomials, where a quadratic term for every ADE polynomial is dropped.

When a $4$d $\mathcal{N}=2$ theory is dimensionally reduced on a circle, at energies lower than the Kaluza-Klein scale, one is left with a $3$d $\mathcal{N}=4$ theory. In particular, we are interested in the dimensional reductions to $3$d of the above-mentioned $(G,G')$ AD theories. Furthermore, for $3$d $\mathcal{N}=4$ theories exists a famous infrared duality named $3$d mirror symmetry \cite{Intriligator:1996ex}. Features of this duality include the fact that the Coulomb and Higgs branches of two dual theories are swapped, as well as the triplets of mass terms and Fayet-Iliopoulos parameters.  It has been a long-standing problem to find the 3d mirror theories for $(G, G')$ theories and there have been a number of papers in the literature that attempted to address this question, \eg~ \cite{Boalch2008,Xie:2012hs, Xie:2013jc, DelZotto:2014kka, Buican:2015hsa, Xie:2017vaf, Benvenuti:2017bpg, Dey:2020hfe, Xie:2021ewm, Closset:2020afy, Dey:2021rxw}. In a series of our previous works \cite{Giacomelli:2020ryy,Carta:2021whq,Carta:2021dyx}, the question of identifying the $3$d mirror theories for AD theories of $(A_n,A_m)$, $(A_n,D_m)$ and $(D_n,D_m)$ types was completely addressed and settled. As a natural follow-up, one of the main results of this paper is the mirror theories for the  $(A_n,E_m)$ type theories for $m=6,7,8$.

The construction of the $3$d mirror theories for AD theories of the $(A_n,E_m)$ type is however not the only result that we propose in this paper. We also utilize higher-form symmetries \cite{Gaiotto:2014kfa} to constrain the dynamics of the theories and obtain certain new results regarding the structure of the Higgs branch and 3d reduction of the $(G,G')$ theories. In fact, higher form symmetries in SCFTs with $8$ supercharges were extensively studied in recent years, \eg~ \cite{DelZotto:2015isa,Albertini:2020mdx,Closset:2020scj,DelZotto:2020esg,Hosseini:2021ged, DelZotto:2022fnw}, and in particular a classification of $1$-form symmetries for all $(G,G')$ AD theories have been achieved recently in \cite{DelZotto:2020esg, Closset:2020scj,Hosseini:2021ged}. In particular, we argue that the $1$-form symmetry of the $4$d AD theory is completely captured in a {\it non-Higgsable sector} that is present at a generic point of the Higgs branch in a large class of AD theories. Such a non-Higgsable sector contains one or several an interacting $4$d SCFTs without any Higgs branch. Via compactification to $3$d, such a sector becomes a collection of twisted hypermultiplets, which upon applying mirror symmetry become free hypermultiplets \cite{Giacomelli:2020ryy}. An immediate consequence of the above statements is that any AD theory with a non-trivial 1-form symmetry gives rise to a free sector upon dimensionally reduction to 3d.

Yet another important result that has proved highly useful in justifying our proposed mirror theories is the proof of the {\it Flip-Flip duality} \cite{Nedelin:2017nsb, Aprile:2018oau} of $T[G]$ theories \cite{Gaiotto:2008ak} for all the simply-laced groups $G$, including the exceptional ones. This duality, along with the action of the Maruyoshi-Song (MS) flow \cite{Maruyoshi:2016tqk}, provides highly non-trivial checks that our proposed $3$d mirror theories for the $(A_n, E_m)$ theories.

The paper is organized as follows. In Section \ref{sec:1formsym} we briefly summarize necessary information about the $1$-form symmetries for $(G,G')$ theories, and discuss their dynamical consequences. In Section \ref{sec:MSproc} we briefly review Flip-Flip duality for $T[G]$ group with $G=\SU(N)$ and then generalize it for any simply-laced group, including $E_{6}$, $E_{7}$ and $E_{8}$. In Section \ref{sec:MirrorAE} we discuss the $3$d mirror theories for the $(A_n,E_m)$ AD theories. The discussion is organized according to the number of mass parameters in descending order. In Section \ref{sec:CommDE} we present certain results on the $(D_n,E_m)$ theories. Finally, in Appendix \ref{app:conv} we introduce our notation and convention of the quivers. In Appendix \ref{app:3dA5E6CBHS}, we describe in detail the computation of the CB Hilbert series for the 3d $\CN=4$ gauge theory resulting from dimensional reduction of the $(A_5, E_6)$ on a circle.  This provides a highly non-trivial test for the proposed mirror theory for the $(A_n, E_6)$ theories.  In Appendix \ref{app:nHSCFTs1form} we propose an updated list of trivially non-Higgsable SCFTs that was given in \cite{Carta:2021whq} with also the corresponding value of electric $1$-form symmetry. Since the non-Higgsable SCFTs of the $(A_n,A_m)$ type always have a trivial $1$-form symmetry, we omit them from this list, but their information can be found in \cite[Table 2]{Carta:2021whq}.

\section{1-form symmetries}
\label{sec:1formsym}
Higher form global symmetries \cite{Gaiotto:2014kfa} has proved useful for constraining the infrared behavior of quantum field theories. In this paper, we use the knowledge of 1-form global symmetries of Argyres-Douglas (AD) theories, studied extensively in \cite{DelZotto:2020esg,Closset:2020scj, Hosseini:2021ged}, to gain the understanding of the structure of their Higgs branch, in particular the non-Higgsable sector that may be present at a generic point, as well as the free sector that arises from dimensional reduction to 3d.  We briefly summarize the necessary information about the 1-form symmetry below.

Let us focus on the AD theories of type $(G,G')$, which arise from Type IIB compactification on isolated hypersurface singularities $\IX_6$ in Type IIB. The lines operators arise from D3-branes wrapping relative 3-cycles inside the desingularization of $\IX_6$. The intersection numbers of these 3-cycles give rise to a defect group that encodes information about the 1-form global symmetry of the theory.  In particular, it can be shown that the electric and magnetic $1$-form symmetries of the $(G, G')$ theory are encoded in the torsional part of the defect group.  The latter takes the form $\mathfrak{f} \oplus \mathfrak{f}$ \cite{DelZotto:2020esg,Closset:2020scj, Hosseini:2021ged} (see also \cite{DelZotto:2015isa,Albertini:2020mdx,Bhardwaj:2021pfz,DelZotto:2022fnw} and \cite{Milnor:1970aaa, Boyer:2007aaa}).  As discussed in \cite{Closset:2020scj, Closset:2021lwy}, it is always possible to choose an electric or a magnetic 1-form symmetry of the theory to be isomorphic to $\mathfrak{f}$. For convenience, we shall henceforth refer to $\mathfrak{f}$ simply as a 1-form symmetry.  Moreover, it was found \cite{Hosseini:2021ged} that the 1-form symmetry is invariant under the Maruyoshi-Song flow; as a consequence, the 1-form symmetry of the $D_p(G)$ theory is the same as that of the $(A_{p-1}, G)$ theory.

Let us now focus on the Higgs branch flow.  One of the main observations in \cite{Giacomelli:2020ryy,Carta:2021whq,Carta:2021dyx} was that there are a number of AD theories that, upon a Higgs branch flow, become a collection of hypermultiplets, plus certain non-trivial SCFTs without Higgs branch.  The latter are referred to as a {\it non-Higgsable sector} or {\it non-Higgsable SCFTs}. Moreover, it was also proposed that reduction to 3d of such a theory yields a collection of twisted hypermultiplets, whose number is equal to the rank of the non-Higgsable SCFT in question. 

With regard to the 1-form symmetry, we make the following observation\\

\noindent\fbox{
\begin{minipage}{0.968\textwidth}
Any AD theory with a non-trivial $1$-form symmetry always has a non-Higgsable sector that has the same $1$-form symmetry as the original theory.
\end{minipage}
}\\

Moreover, we find that if an AD theory has a trivial 1-form symmetry, then either it does not have a non-Higgsable sector or the non-Higgsable sector possesses a trivial 1-form symmetry.
In other words, the 1-form symmetry is ``trapped'' in the non-Higgsable sector. 
An immediate dynamical consequence of this statement is that a theory resulting from dimensional reduction of an AD theory with a non-trivial $1$-form symmetry to 3d always contains a free sector, namely $r\geq 1$ twisted hypermultiplets, where $r$ is the total rank of the SCFTs in the corresponding non-Higgsable sector. 

An implication of the above statement is that the 1-form symmetry is invariant under the Higgs branch flow. This should be contrasted with the Coulomb branch flow, where the BPS states may screen the 1-form symmetry \cite{DelZotto:2022fnw}.

Subsequently, we will use the above observation to constrain the properties of the Higgs branch of several AD theories with non-trivial 1-form symmetries. For $(A_n, D_m)$ and $(D_n, D_m)$ types theories, such an observation put constraints on the non-Higgsable SCFTs that are in perfect agreement with the findings of \cite{Carta:2021whq, Carta:2021dyx}. For $(A_n, E_m)$ theories, the above observation turns out to be very useful as a guiding principle to identify the non-Higgsable SCFT and hence the mirror theory. For reference, we provide \cref{tab:AnDmNHC,tab:GGpNHC}, containing a list of non-Higgsable SCFTs of the type $(G,G')$ with $0 \leq 24(c-a) < 1$, together with their ranks and 1-form symmetries.  Note that the $(A_n, A_m)$ theories are excluded there, since all of them have a trivial 1-form symmetry \cite{DelZotto:2020esg,Closset:2020scj} and their information has already been contained in \cite[Appendix C]{Carta:2021whq}.

\subsection{Realization in a weakly-gauged description}
\label{sec:1formonCM}
It was shown in \cite{Buican:2021xhs} that any Argyres-Douglas theory of type $(G,G')$ that does not admit an $\CN=2$ preserving marginal deformation cannot have a non-trivial 1-form symmetry. In this section, we thus focus on the theories with a $\CN=2$ preserving conformal manifold and discuss the realization of the $1$-form symmetry at a weak-coupling cusp.

For simplicity, we focus on the $(D_n,D_m)$ theories,\footnote{We will discuss the $(A_n, A_m)$ theories below.  For this class of theories, it is well-known the 1-form symmetry of these theories are trivial \cite{DelZotto:2020esg, Closset:2020scj}. For $(A_n, D_m)$, the discussion in this section can also be applied using information from \cite{Carta:2021whq, Buican:2021xhs}, although the details are rather lengthy.  For $(A_n, E_{m})$, $(D_n, E_{m})$ and $(E_n, E_m)$, we currently do not have a weakly-gauged description. We leave the latter cases for future work.} whose conformal manifolds were extensively studied in \cite{Carta:2021dyx}.\footnote{See also \cite{Buican:2014hfa} for a discussion about conformal manifolds of Argyres-Douglas theories.}  In fact, using the information from \cite{Carta:2021dyx}, along with \cite{Closset:2020scj, DelZotto:2020esg, Hosseini:2021ged}, we observe that 
\ben
\item a $(D_n,D_m)$ theory that has a $\CN=2$ preserving non-trivial conformal manifold admits either one or two or six mass parameters, and
\item those with one or two mass parameters have non-trivial $1$-form symmetry, while those with $6$ mass parameters always have a trivial $1$-form symmetry. \label{oneformDnDm}
\een
In the following, we will explain why observation \ref{oneformDnDm} holds. 

As shown in \cite{Carta:2021dyx}, any $(D_n,D_m)$ theory that admit a $\CN=2$ preserving non-trivial manifold can be realized as gauging of two $D_p(\SO(\text{even}))$ theories via a special orthogonal gauge algebra. Below, we will demonstrate that the global form of such a gauge algebra should be a $\Spin$ gauge group. In which case, we write
\begin{equation} \label{glueDnDm}
    (D_n,D_m): \qquad D_{p_1}(\SO(2N_1)) - \Spin(2N_2) - D_{p_2} (\SO(2N_2))\coma
\end{equation}
assuming that $\mathfrak{so}(2N_2)$ is a subalgebra of $\mathfrak{so}(2N_1)$, i.e. $N_1>N_2$. The $1$-form symmetry of the $(D_n, D_m)$ theory in \eref{glueDnDm} is then the product of the following symmetries:
\begin{enumerate}[label={(\arabic*)},ref={(\arabic*)}]
    \item the $1$-form symmetry of $D_{p_1}(\SO(2N_1))$, \label{list:sym1}
    \item the $1$-form symmetry of $D_{p_2}(\SO(2N_2))$, and  \label{list:sym2}
    \item the center symmetry of $\Spin(2N_2)$ that is not screened by the Higgs branch generators of $D_{p_1}(\SO(2N_1))$ and $D_{p_2}(\SO(2N_2))$.  \label{list:sym3}
\end{enumerate}
The symmetries in \ref{list:sym1} and \ref{list:sym2} can be determined from the fact that the $1$-form symmetry is invariant under the Maruyoshi-Song flow \cite{Hosseini:2021ged}, and so they are exactly the same as the $1$-form symmetries of  $(A_{p_1-1}, D_{N_1})$ and $(A_{p_2-1}, D_{N_2})$ respectively. The latter are given in \cite{Closset:2020scj, DelZotto:2020esg, Hosseini:2021ged}.  To complete the discussion, we now discuss \ref{list:sym3}.  Some ideas below will be similar to that of \cite{Bhardwaj:2021mzl}.

Let us state certain important observations that we will soon need.
\bi
\item For $p$ odd and $p\geq 2N-2$, the Higgs branch of $D_{p}(\SO(2N))$ is generated by the moment maps in the adjoint representation of $\mathfrak{so}(2N)$  This can be seen as follows.  The mirror theory for such a $D_{p}(\SO(2N))$ theory is given by the $T[\SO(2N)]$ theory plus a collection of free hypermultiplets \cite[(5.4)]{Carta:2021whq}. Since the Higgs/Coulomb branch of $T[\SO(2N)]$ is generated by the moment map in the adjoint representation of $\mathfrak{so}(2N)$, so is the Higgs branch of $D_{p}(\SO(2N))$.

For $p$ odd and $p< 2N-2$, the 3d reduction is $T_\sigma[\SO(2N)]$ for some partition $\sigma$ (see \cite[(4.2)]{Carta:2021dyx}). Since the Higgs branch of this theory is generated by the moment map in the adjoint representation of $\mathfrak{so}(2N)$, the same statement also holds for the Higgs branch of the corresponding $D_{p}(\SO(2N))$ theory.
\item Let us now take $p$ to be even and consider the $D_p(\SO(2N))$ theories with $(2N-2)/\GCD(2N-2,p)$ odd.  The mirror for $p> 2N-2$ and $p< 2N-2$ are given in \cite[Section 6.1.2]{Carta:2021whq} and  \cite[Section 4.2.2]{Carta:2021dyx}, respectively.  In either case, there is an overall $\ZZ_2$ that has to be quotiented out in quiver description of the mirror theory. Upon computing the Coulomb branch generators, the half-integral value fluxes give rise to those in the spinor representation of $\Spin(2N)$, whereas the integral fluxes give rise to those in the adjoint representation of $\Spin(2N)$.
\item Let us now take $p$ to be even and consider the $D_p(\SO(2N))$ theories with $(2N-2)/\GCD(2N-2,p)$ even.  For $p\geq 2N-2$ its 3d reduction gives rise to the quiver given by \cite[(6.6)]{Carta:2021whq}, whereas for $p<2N-2$ its 3d reduction is discussed in \cite[Section 4.2.1]{Carta:2021dyx}.  In either case, the interacting part of the 3d reduction is described in the following quiver theory\footnote{The global symmetry $D_N$ and $D_1$ can be explicitly seen from the partition $\rho= [(p-1)^2, 1^{2N}]$ in \cite[Section 4.2.1]{Carta:2021dyx}.}
\bes{
[D_{N}]- \cdots -[D_1]
}
where $\cdots$ denotes alternating $C$-type and $D$-type gauge groups. The Higgs branch generators of this theory and therefore of the 4d theory consist of the meson in the adjoint representation of $\mathfrak{so}(2N)$, the singlet of $\mathfrak{so}(2)$, and the ``long meson'' in the bifundamental representation of $\mathfrak{so}(2N)\times \mathfrak{so}(2)$.

\ei
Let us now apply these points to \eref{glueDnDm} as follows.
\begin{enumerate}
    \item \label{point1}  For the $(D_n, D_m)$ theories with one mass parameter, we observe that both $p_1$ and $p_2$ are odd.  As we pointed out above, the Higgs branches of $D_{p_1}(\SO(2N_1))$ and $D_{p_2}(\SO(2N_2))$ are generated by the moment maps in the adjoint representations of $\mathfrak{so}(2N_1)$ and $\mathfrak{so}(2N_2)$, respectively.  The decomposition of the adjoint representation of $\mathfrak{so}(2N_1)$ into the representation of $\mathfrak{so}(2N_2)$ always contains the adjoint representation and vector representation, but not a spinor representation. The vector representation screens a $\ZZ_2$ subgroup in the $\ZZ_2\times \ZZ_2$ or $\ZZ_4$ center of $\Spin(2N_2)$.  Hence, we always have a $\ZZ_2$ factor in the $1$-form symmetry coming from the center of $\Spin(2N_2)$ gauge group in \eref{glueDnDm} that is not screened.  As a consequence, the $1$-form symmetry for $(D_n,D_m)$ with one mass parameter can never be trivial.
    \item For the $(D_n, D_m)$ theories with two mass parameters, one of $p_1$ and $p_2$ is odd and the other is even.  For the even one (let us denote it by $p$), the corresponding $D_p(\SO(2N))$ has $(2N-2)/\GCD(2N-2,p)$ being even. The ``long meson'' in the bifundamental of $\mathfrak{so}(2N)\times \mathfrak{so}(2)$ gives rise to the quantities in the vector representation of $\mathfrak{so}(2N)$ that screens a $\ZZ_2$ subgroup in the $\ZZ_2\times \ZZ_2$ or $\ZZ_4$ center of $\Spin(2N_2)$. There is no Higgs branch generator in the spinor representation from either $D_{p_1}(\SO(2N_1))$ or $D_{p_2}(\SO(2N_2))$.  Hence, we always have a $\ZZ_2$ factor in the $1$-form symmetry coming from the center of $\Spin(2N_2)$ gauge group in \eref{glueDnDm} that is not screened.  As a consequence, the $1$-form symmetry for $(D_n,D_m)$ with two mass parameters can never be trivial.
    \item For the $(D_n, D_m)$ theories with six mass parameters, both $p_1$ and $p_2$ are even. Both theories are always of the form $D_p(\SO(2N))$ with even $p$ such that $(2N-2)/\GCD(2N-2,p)$ is odd. As we mentioned above, each theory has Higgs branch generators in the spinor representation and adjoint representation.  The former screens one $\ZZ_2$ factor in the center symmetry of $\Spin(2N_2)$, and the decomposition of the latter of $\mathfrak{so}(2N_1)$ to the subgroup $\mathfrak{so}(2N_2)$ gives rise to the vector representation that screens the other $\ZZ_2$ factor in the center symmetry (see Point \ref{point1}).  In summary, the $\ZZ_2\times\ZZ_2$ or the $\ZZ_4$ center of $\Spin(2N_2)$ is totally screened.  Next, we show that in this case, both $D_{p_1}(\SO(2N_1))$ and $D_{p_2}(\SO(2N_2))$ have a trivial $1$-form symmetry.  We can perform a Maruyoshi-Song flow from the $D_p(\SO(2N))$ theory to the $(A_{p-1}, D_{N})$ theory.  Using the fact that the $1$-form symmetry is invariant under the Maruyoshi-Song flow and that the $(A_{p-1}, D_N)$ theory with even $p$ and $(2N-2)/\GCD(2N-2,p)$ odd has a trivial $1$-form symmetry\footnote{This is because $2$ divides $p$ and $\GCD(p, 2N-2)$ does not divide $N-1$, since $(2N-2)/\GCD(2N-2,p)$ is odd; see Table 1 of \cite{Hosseini:2021ged}.} \cite{Hosseini:2021ged}, we reach the conclusion that the corresponding $D_p(\SO(2N))$ theory also has a trivial $1$-form symmetry. As a consequence, the $1$-form symmetry for $(D_n,D_m)$ with six mass parameters is always trivial. 
\end{enumerate}

We can apply the same argument to the $(A_n, A_m)$ theories with a non-trivial conformal manifold. As discussed in \cite[(3.26)]{Giacomelli:2020ryy}, such a theory can be written as a $\SU(N_2)$ gauging of certain $D_{p_1}(\SU(N_1))$ and $D_{p_2}(\SU(N_2))$ with $N_1 >N_2$ and $p_2>N_2$. Note that a $D_p(\SU(N))$ theory has a trivial $1$-form symmetry.\footnote{The reason for this is as follows. There is a Maruyoshi-Song (MS) flow from the $D_p(\SU(N))$ theory to the $(A_{p-1}, A_{N-1})$ theory \cite{Maruyoshi:2016aim, Agarwal:2016pjo}. Due to the fact that the 1-form symmetry of the latter theory is trivial \cite{DelZotto:2020esg, Closset:2020scj} and that the 1-form symmetry is invariant under the MS flow \cite{Hosseini:2021ged}, we conclude that the 1-form symmetry of the $D_p(\SU(N))$ theory is trivial.} Since the branching rule of the adjoint representation, under which the moment map transforms, of $\SU(N_1)$ to $\SU(N_2)$ always contains a fundamental and an antifundamental representations of $\SU(N_2)$, these screen the $\BZ_{N_2}$ center of the $\SU(N_2)$ gauge group. Overall, the $1$-form symmetry is trivial, as expected for the $(A_n, A_m)$ theories.

As a matter of fact, this analysis can be applied to any theory involving weakly gauging of SCFTs. We demonstrate this in a subsequent example \eref{A8E7}.    

\section{Maruyoshi-Song procedure}
\label{sec:MSproc}

In order to extract the 3d mirrors of $(A,E)$ theories, we can exploit the fact that all $(A,G)$ models have a class $\mathcal{S}$ description involving a sphere with a single irregular puncture for the 6d $\mathcal{N}=(2,0)$ theory of type $G$. If we add to this setup a full regular puncture we find the so-called $D_p(G)$ theories and from these we can flow back to $(A_{p-h^{\vee}_G-1},G)$, for $p\geq h^{\vee}_G$,\footnote{The constraint comes from the fact that chiral ring relations prevent us from activating a principal nilpotent VEV for the moment map for low values of $p$.} (where $h^{\vee}_G$ denotes the dual Coxeter number of $G$) by activating a principal nilpotent VEV for the $G$ moment map. As was observed in \cite{Maruyoshi:2016aim, Agarwal:2016pjo, Giacomelli:2017ckh}, we can also flow from $D_p(G)$ to $(A_{p-1},G)$ (this time without restrictions on the value of $p$) with a variant of the above procedure (MS flow), which involves coupling a chiral multiplet to the $G$ moment map via a superpotential term and activating a nilpotent VEV for it, instead of the moment map. This generalizes a class of RG flows discovered in \cite{Maruyoshi:2016tqk}. We will refer to these adjoint chirals as flipping fields from now on. 

This observation is relevant for us because we can exploit it to relate the mirrors of $(A_{p-1},G)$ and $(A_{p-h^{\vee}_G-1},G)$ and therefore constrain the problem. In order to do this, we first have to discuss a 3d duality we will need below. 

\subsection{Flip-Flip duality for $T[G]$} 

The 3d $\mathcal{N}=4$ theory $T[G]$ introduced in \cite{Gaiotto:2008ak} can be thought of as the S-duality wall for $\mathcal{N}=4$ super Yang-Mills with gauge group $G$ and is used to construct the S-dual of the one-half BPS Dirichlet boundary condition for the same theory: One simply couples the fields of $\mathcal{N}=4$ SYM to $T[G]$ at the boundary. The theory has global symmetry $G$ on the Higgs branch and $G^{\vee}$ on the Coulomb branch, where $G^{\vee}$ denotes the langlands dual group. The mirror dual of $T[G]$ is known to be $T[G^{\vee}]$. In particular $T[G]$ is self-mirror for $G$ simply-laced, which is the only case we will need in this paper. 

For $G$ classical and simply-laced the $T[G]$ theory has a known quiver description and also a Hanany-Witten brane realization in Type IIB string theory. The explicit Lagrangian description was exploited in \cite{Nedelin:2017nsb} to prove that $T[\SU(N)]$ has a dual description (dubbed Flip-Flip duality in \cite{Nedelin:2017nsb}) analogous to Aharony duality \cite{Aharony:1997gp} for $\mathcal{N}=2$ SQCD. The dual description simply involves flipping the moment maps both on the Higgs branch and on the Coulomb branch, where the flipping fields are the chiral ring counterpart of the corresponding moment maps in the original theory. Notice that this operation naively breaks extended supersymmetry to $\mathcal{N}=2$ but it turns out supersymmetry enhances back to $\mathcal{N}=4$ in the infrared, as the duality implies. The analogy with Aharony duality is due to the fact that the Flip-Flip duality can be derived by performing a sequence of Aharony dualities on the gauge nodes of the quiver. This, in particular, allows one to check the supersymmetry enhancement mentioned before. 

The Flip-Flip duality can also be given a brane interpretation, as flipping moment maps turns out to be equivalent to rotating the corresponding 5-branes. Flipping both the Higgs branch and Coulomb branch moment maps actually corresponds to rotating the brane system as a whole, and this is a trivial operation which does not affect the field theory. This argument carries over to the $G=\SO(2N)$ case, as pointed out in \cite{Carta:2021whq} and therefore one might suspect this holds also for $G$ exceptional, even though the arguments presented above do not apply since neither a brane realization nor a Lagrangian description are known. We are now going to present an argument which applies to every $G$ simply-laced and only uses the properties of BPS boundary conditions for $\mathcal{N}=4$ SYM with gauge group $G$.

First, we describe the $T[G]$ theory as the infrared limit of $\mathcal{N}=4$ $G$ SYM on an interval, coupled to $T[G]$ at one end of the interval, say on the left, and with Dirichlet boundary conditions at the right end (see \cite{Gaiotto:2008ak}). In order to discuss more in detail the boundary conditions, we split the six adjoint scalars in the 4d $\mathcal{N}=4$ multiplet into two groups of three: $X_{1,2,3}$ and $Y_{1,2,3}$. We also denote their boundary values on the left ($L$) and right ($R$) of the interval by $X^{L/R}_{1,2,3}$ and $Y^{L/R}_{1,2,3}$. The 4d theory on the interval described above preserves eight supercharges if we give Dirichlet boundary conditions on the right to three of the scalars (say $Y_i^R$) and Neumann boundary conditions on the right to the other three (namely $X_i^R$) \cite{Gaiotto:2008sa}. In order to make contact with the brane description available for $G$ classical, we can identify $X_i^{R}$ with the directions wrapped by the D5-branes\footnote{The D5-branes impose the Dirichlet boundary condition on the gauge fields and on $Y^R_{1,2,3}$, parametrizing the transverse directions to the D3-branes.} in Type IIB. At the other end of the interval the scalars $Y_i^L$ are given Neumann boundary conditions whereas the $X_i^L$ are forced to be equal to the Higgs branch moment map of $T[G]$ due to the coupling with the 3d theory. In order to proceed with our argument, it is convenient to combine $X_{1,2}^{L/R}$ into a complex field $X_{\mathbb{C}}^{L/R}\equiv X_1^{L/R}+iX_2^{L/R}$ and similarly introduce $Y_{\mathbb{C}}^{L/R}\equiv Y_1^{L/R}+iY_2^{L/R}$. 

We now modify the above setup by interchanging $X_{\mathbb{C}}^R$ and $Y_{\mathbb{C}}^R$: We give Neumann boundary conditions to $Y_{\mathbb{C}}^R$ (which used to be Dirichlet) and Dirichlet to $X_{\mathbb{C}}^R$ (which used to be Neumann), 
whereas we leave the boundary condition on the left unchanged. With this modification the system only preserves four supercharges and not eight as before \cite{Hashimoto:2014vpa} and corresponds for classical $G$ to rotating the D5-branes. Since now $X_{\mathbb{C}}^R$ has Dirichlet boundary conditions, due to the BPS equation, supersymmetry prevents $X_{\mathbb{C}}$ to have a non-trivial dependence on the coordinate along the interval.  This implies that $X_{\mathbb{C}}^L = X_{\mathbb{C}}^R$. As a result, the complex part of the $T[G]$ Higgs branch moment map is set to zero, which is exactly the effect we expect from the flipping. Moreover, since $Y_{\mathbb{C}}^{L}=Y_{\mathbb{C}}^{R}$ now has Neumann boundary conditions, 
its value is unconstrained and can be identified with the VEV of the flipping field  (see \cite{Gaiotto:2015una} for a similar construction). By interchanging the Dirichlet and Neumann boundary conditions on the right, we interchange the roles of $X_{\mathbb{C}}^R$ and $Y_{\mathbb{C}}^R$.  This is equivalent to flipping the Higgs branch moment map of $T[G]$. 

In order to conclude the argument, we have to exploit the self-mirror property of $T[G]$ for $G$ simply laced. Performing S-duality for the 4d theory in the bulk of the interval interchanges the 3d theory with its mirror and turns the boundary condition on the left into Dirichlet boundary condition. Likewise, the boundary condition on the right becomes the S-dual of Dirichlet, namely the coupling to $T[G^{\vee}]=T[G]$. Since we know that the mirror of $T[G]$ with the Higgs branch moment map flipped is $T[G]$ with the Coulomb branch moment map flipped, we learn that interchanging the roles of $X_{\mathbb{C}}^L$ and $Y_{\mathbb{C}}^L$, 
where we had the S-dual of Dirichlet, corresponds to flipping the Coulomb branch moment map. At this stage, we conclude just by observing that flipping both moment maps corresponds to interchanging $X_{\mathbb{C}}^{L}$ and $Y_{\mathbb{C}}^{L}$ as well as $X_{\mathbb{C}}^{R}$ and $Y_{\mathbb{C}}^{R}$, 
but this is a simple relabeling of the scalar fields which cannot have any effect on the infrared theory. We therefore obtain the Flip-Flip duality for any $G$ simply-laced.  

\subsection{$D_p(G)$ theories and special unitary quivers with exceptional shape}\label{minpuncture}

In order to proceed with our analysis, we find it convenient to work out in detail the properties of $D_p(G)$ theories with $G$ simply-laced and $p=h^{\vee}_G$, the dual Coxeter number of $G$. We are specifically interested in understanding the Higgs branch of these models. In order to solve this problem we follow a slightly indirect route, which involves partially closing the full $G$ puncture in the class $\mathcal{S}$ description of the theory to a minimal puncture\footnote{Note that for an exceptional gauge group $E_n$, the full puncture is denoted by the Bala-Carter label $0$ and the minimal puncture is denoted by $E_n(a_1)$.} and then proceed by analyzing the resulting model. The advantage of this approach is that, as we will see, the theory with a minimal puncture turns out to be Lagrangian, which indeed greatly simplifies the analysis, and more specifically it is described by a special unitary quiver with affine $G$ shape. In particular, for $G$ exceptional of the $E_{6,7,8}$-type, we find {\it for the first time} the class $\mathcal{S}$ realization of quivers with the shape of the corresponding {\it exceptional affine Dynkin diagram} such that all gauge groups are {\it special unitary}. 

Let us directly focus on the case of $G$ exceptional of the $E_{6, 7,8}$-type. The fact that $p=h^\vee_G$ implies that the CB spectrum of these theories is given by
\begin{equation}
    \text{CB} = \bigcup_{\text{C}(E_n)}\left\{i = 2,\ldots, \text{C}(E_n)-1 \right\}\coma
    \label{eq:CBDpEn}
\end{equation}
where, by $\text{C}(E_n)$, we mean the Casimir operators of $E_n$, with $n=6,7,8$. Let us consider, for instance, $D_{12}(E_6)$. The CB spectrum of this theory according to \eqref{eq:CBDpEn} is 
\begin{equation}
\begin{split}
    \{&\{\},\{2,3,4\},\{2,3,4,5\},\{2,3,4,5,6,7\},\{2,3,4,5,6,7,8\},\\
    & \{2,3,4,5,6,7,8,9,10,11\}\}\coma
\end{split}
    \label{eq:CBDpE6}
\end{equation} 
while, by the classification of punctures of type $E_6$ in \cite{Chacaltana:2014jba}, the pole structure of the full puncture in type $E_6$ theories is
\begin{equation}
    p^{\max}_{E_6}=\left\{p^{\max}_2,p^{\max}_5,p^{\max}_6,p^{\max}_8,p^{\max}_9,p^{\max}_{12}\right\} = \{1,4,5,7,8,11\}\fstop
\end{equation}
Since there are no constraints on the CB operators for the maximal puncture, the pole structure  counts the number of masses (one for each Casimir) plus the CB operators in \eqref{eq:CBDpE6}. 
The procedure now involves replacing the full puncture of $D_{h^\vee_E}(E)$ by the minimal puncture of the corresponding exceptional group $E=E_{6,7,8}$. The resulting theory has a spectrum compatible with a quiver shaped like the affine Dynkin diagram of $E$. This is obtained by removing from the spectrum of $D_{h^\vee_E}(E)$ the operators which are not compatible with the pole structure and constraints of the minimal puncture. Let us again consider $D_{12}(E_6)$ where we replace the full puncture by a minimal one, whose pole structure is \cite{Chacaltana:2014jba}
\begin{equation}
    p^{\min}_{E_6}=\left\{p^{\min}_2,p^{\min}_5,p^{\min}_6,p^{\min}_8,p^{\min}_9,p^{\min}_{12}\right\} = \{1,1,2,2,2,3\}\fstop
\end{equation}
For this theory, the minimal puncture does not have any constraint, so the resulting spectrum can be obtained by removing from the original CB spectrum a number of operators given by the difference between the pole structure of the full puncture and the pole structure of the minimal puncture,\footnote{Generically, one needs to obtain the graded CB spectrum from the pole structure by imposing the constraints coming from the analysis of the Hitchin system.} i.e.
\begin{equation}
    p^{\max}_{E_6} - p^{\min}_{E_6} = \{0,3,3,5,6,8\}\fstop 
\label{eq:pmaxpminE6}
\end{equation}
To determined the CB spectrum, we remove the elements, in ascending order, in each list in \eqref{eq:CBDpE6} so that each list has the number of the elements equal to \eqref{eq:pmaxpminE6}. The elements that are removed constitute the CB spectrum of the theory with the minimal puncture. Below, we highlight such elements in \eqref{eq:CBDpE6} in red:
\bes{
\{& \{\},\{2,3,4\},\{{\red 2},3,4,5\},\{{ \red 2},3 ,4,5,6,7\},\{{\red 2}, 3,4,5,6,7,8\},\\
&\{{\red 2,3},4,5,6,7,8,9,10,11\}\}~.
}
The resulting CB spectrum is thus determined by the red elements: 
\begin{equation}
\text{CB}^{\min}(E_6) = \{2,2,2,2,3\}~.
\end{equation}
This is compatible with that of the following quiver:
\bes{ \label{eq:E6minpunc}
\begin{tikzpicture}[baseline,font=\footnotesize]
\node[draw=none] (SU1l) at (-4,0) {$\SU(1)$};
\node[draw=none] (SU2l) at (-2,0) {$\SU(2)$};
\node[draw=none] (SU3) at (0,0) {$\SU(3)$};
\node[draw=none] (SU2u) at (0,1) {$\SU(2)$};
\node[draw=none] (SU1u) at (0,2) {$\SU(1)$};
\node[draw=none] (SU2r) at (2,0) {$\SU(2)$};
\node[draw=none] (SU1r) at (4,0) {$\SU(1)$};
\draw (SU1l)--(SU2l)--(SU3)--(SU2r)--(SU1r);
\draw (SU3)--(SU2u) -- (SU1u);
\end{tikzpicture}
}
where $\SU(1)$ should be viewed as one flavor attached to the gauge node next to it.  Thus, by closing the full $E_6$ puncture in $D_{12}(E_6)$ to the minimal $E_6$ puncture, we obtain an affine $E_6$-shaped quiver with all nodes being special unitary, as promised.  We remark that \eref{eq:E6minpunc} is in fact the Lagrangian description of the $(D_4, D_4)$ theory; see \cite[(6.6)]{Closset:2020afy} and \cite[(5.53)]{Carta:2021dyx}.

One can repeat the same discussion for $D_{18}(E_7)$, with the following data \cite{Chacaltana:2017boe}
\begin{equation}
    \begin{split}
      p^{\max}_{E_7}&=\left\{p^{\max}_2,p^{\max}_6,p^{\max}_8,p^{\max}_{10},p^{\max}_{12},p^{\max}_{14},p^{\max}_{18}\right\} = \{1,5,7,9,11,13,17\}\coma  \\
      p^{\min}_{E_7}&=\left\{p^{\min}_2,p^{\min}_6,p^{\min}_8,p^{\min}_{10},p^{\min}_{12},p^{\min}_{14},p^{\min}_{18}\right\} = \{1,2,2,2,3,3,4\} \coma
    \end{split}
\end{equation}
obtaining, by subtraction analogously to the $D_{12}(E_6)$ case, the following CB spectrum
\begin{equation}
    \mathrm{CB}^{\min}(E_7) = \{2,2,2,2,2,2,3,3,3,4\}\fstop
\end{equation}
The resulting theory is compatible with the following quiver
\bes{ \label{eq:E7minpunc}
\begin{tikzpicture}[baseline,font=\footnotesize]
\node[draw=none] (SU1l) at (-6,0) {$\SU(1)$};
\node[draw=none] (SU2l) at (-4,0) {$\SU(2)$};
\node[draw=none] (SU3l) at (-2,0) {$\SU(3)$};
\node[draw=none] (SU4) at (0,0) {$\SU(4)$};
\node[draw=none] (SU2u) at (0,1) {$\SU(2)$};
\node[draw=none] (SU3r) at (2,0) {$\SU(3)$};
\node[draw=none] (SU2r) at (4,0) {$\SU(2)$};
\node[draw=none] (SU1r) at (6,0) {$\SU(1)$};
\draw (SU1l)--(SU2l)--(SU3l)--(SU4)--(SU3r)--(SU2r)--(SU1r);
\draw (SU4)--(SU2u);
\end{tikzpicture}
}
Thus, by closing the full $E_7$ puncture in $D_{18}(E_7)$ to the minimal $E_7$ puncture, we obtain an affine $E_7$-shaped quiver with all nodes being special unitary, as promised.

The theory $D_{30}(E_8)$ also follows using the pole structures in \cite{Chacaltana:2018vhp}, however, it seems that the minimal puncture contains non-trivial constraints to be taken into account and these are not available in the literature yet. We can however notice that, as in the $E_6$ and $E_7$ cases, the central charges $a$ and $c$ are compatible with an $E_8$-shaped quiver. 

We can, of course, repeat the analysis also for $G$ classical. Indeed, for $G=\SU(N)$ the theory with a minimal puncture describes a collection of $N$ free hypermultiplets whereas for $G=\SO(2N)$ we find a linear quiver with $N-3$ $\SU(2)$ gauge groups, terminating with two flavors on both sides. 

The important fact for us is that for every choice of $G$ the Higgs branch of the minimal puncture theory contains as a subvariety the Kleinian singularity $\mathbb{C}^2/\Gamma_G$, where $\Gamma_G$ is the discrete group corresponding to $G$, and indeed it becomes isomorphic to it upon gauging all the baryonic $\U(1)$ symmetries of the theory, whose number is equal to the rank of $G$. Moving along this subvariety physically corresponds to closing the minimal puncture. We therefore learn that the full puncture of $D_{h^{\vee}_G}(G)$ can be fully closed. 

\subsection{The mirror dual of $D_{h^{\vee}_G}(G)$ and $(A_{\fn h^{\vee}_G-1},G)$}   \label{sec:genprophG}

Let us now apply our findings to the determination of the 3d mirrors of Argyres-Douglas theories. In the following, we denote by $r_G$ the rank of $G$.  

We start by looking at the properties of $D_{h^{ \vee}_G}(G)$:
\begin{itemize}
\item Its $G$ moment map is not constrained by chiral ring relations and can be given a principal nilpotent VEV. We therefore expect the Higgs branch to include as a subvariety the $G$ principal nilpotent orbit, whose quaternionic dimension is $h^{\vee}_Gr_G/2$. 
\item The dimension of the Higgs branch is $h^{\vee}_Gr_G/2+r_G$ and we do not expect any non-Higgsable sectors. 
\item The number of mass parameters is $2r_G$. 
\item The Coulomb branch dimension is $h^{\vee}_Gr_G/2-r_G$. 
\end{itemize}
The first two statement follow from the properties of the minimal puncture theories discussed in Section \ref{minpuncture} and the last two directly follow from the results of \cite{Cecotti:2012jx, Cecotti:2013lda} and \cite{Giacomelli:2017ckh}.
All these properties are compatible with a 3d mirror theory given by $T[G]$ with the Cartan of the Higgs branch symmetry gauged. Our claim is therefore that the 3d mirror of $D_{h^{\vee}_G}(G)$ is just 
\be T[G]///\U(1)^{r_G}~,\ee
where $///$ denotes gauging (hyperK\"ahler quotient at the level of the moduli space). We indeed know this is the case for $G$ classical, and our findings strongly suggest this holds for exceptional groups as well.  
We can now exploit the fact that with the MS flow we can go from $D_{h^{\vee}_G}(G)$ to $(A_{h^{\vee}_G-1},G)$. Coupling the flipping field to the $G$ moment map of the 4d theory amounts on the 3d mirror side to flipping the Coulomb branch moment map of $T[G]$ and thanks to the Flip-Flip duality this is equivalent to flipping the Higgs branch. Exploiting this, it is now a simple task to understand the effect of the RG flow by applying the methods developed in \cite{Benvenuti:2017lle, Benvenuti:2017kud, Benvenuti:2017bpg}, since turning on the nilpotent VEV just amount to removing $T[G]$ from the 3d theory. We therefore end up with a $\U(1)^{r_G}$ abelian theory with matter fields organized into hypermultiplets which are in one-to-one correspondence with the positive roots of $G$. We will describe these field theories more in detail in the next section. 

Moreover, we can exploit the MS flow to establish a sort of ``recursion relation" among 3d mirrors of $(A,G)$ theories: Once we know the mirror of $(A_{p-1},G)$, we also know that the 3d mirror of $D_{p+h^{\vee}_G}(G)$ is the same theory coupled to $T[G]$ via a gauging. With an MS flow starting from the latter, we can then obtain the 3d mirror of $(A_{p+h^{\vee}_G-1},G)$ in which the $T[G]$ theory is removed and traded for a chiral in the adjoint of $G$, which contributes $h^{\vee}_Gr_G/2$ hypermultiplets in one-to-one correspondence with the positive roots of $G$. 

Let us now generalize the above observation to the $\left(A_{\fn h^\vee_G-1} ,G\right)$ theories. The number of mass parameters is $r_G$. This theory has rank $\left(\frac{1}{2} \fn h^\vee_G-1 \right) r_G$, the value of $24(c-a)$ equal to $r_G$, and a trivial $1$-form symmetry.  In general, we propose that the mirror theory is described by a 3d $\CN=4$ gauge theory with $r_G$
abelian gauge groups and with the Higgs branch dimension $\left(\frac{1}{2} \fn h^\vee_G-1 \right)r_G$, 
\ie~ there are $\frac{1}{2}\fn h^\vee_G r_G$ 
hypermultiplets in this description.  Recall again that $\frac{1}{2} h^\vee_G r_G$ 
is the number of positive roots of $G$. This, along with the above discussion, leads us to propose that the matter content of the mirror theory for $\left(A_{\fn h^\vee_G-1} ,G\right)$ is described by $\fn$ copies of a collection of hypermultiplets, whose charges are given by the positive root vectors of $G$ with integer elements.  There is no free hypermultiplet in the mirror theory.  Let us demonstrate this in explicit examples.

Let $\{\vec e_i\}$
be an orthonormal basis. For $G=A_{m}=\mathfrak{su}(m+1)$, the positive roots are (see \eg~ \cite[Appendix A.1]{Keller:2011ek})
\bes{
\{\vec e_i - \vec e_j\}_{1 \leq i < j \leq m+1}\fstop
}
The mirror theory for $(A_{\fn (m+1) -1}, A_{m})$ is therefore described by a 3d $\CN=4$ gauge theory with $(m+1)$ $\U(1)$ gauge nodes and the $\fn$ copies of a collection of hypermultiplets, each with charge $(+1, -1)$ under the $i$-th and $j$-th gauge groups such that $1\leq i <j \leq m+1$. This can be conveniently described by a quiver diagram that is a complete graph with $(m+1)$ $\U(1)$ nodes, where each edge has multiplicity $\fn$. This is precisely in agreement with the description presented below \cite[(4.12)]{Giacomelli:2020ryy}. It should be emphasized that there is an overall $\U(1)$ that decouples, and so the number of abelian group that acts non-trivially on the hypermultiplet is actually $\rank\, A_m = m$, in accordance with the above discussion.

For $G=D_m=\mathfrak{so}(2m)$, the positive roots are (see \eg~ \cite[Appendix A.1]{Keller:2011ek})
\bes{
\{ \vec e_i + \vec e_j \}_{1 \leq i <j \leq m } \cup \{ \vec e_i - \vec e_j \}_{1 \leq i <j \leq m } \fstop
}
The mirror theory for $(A_{\fn (2m-2)-1}, D_m)$ is therefore described by a 3d $\CN=4$ gauge theory with $m$ $\U(1)$ gauge nodes and the $\fn$ copies of a collection of hypermultiplets coming in two sets: One with charge $(+1, +1)$ and the other with charge $(+1, -1)$ under the $i$-th and $j$-th gauge group such that $1 \leq i <j \leq m$.  Since a pair of hypermultiplets with charges $(+1, +1)$ and $(+1, -1)$ under $\U(1) \times \U(1)$ can be replaced by four half-hypermultiplets in the representation $[\mathbf{2},\mathbf{2}]$ of $\SO(2) \times \SO(2)$, see \eref{blueedge} and \eref{repblueedge}, this description can be conveniently rephrased in terms of a quiver diagram that is a complete graph of $m$ $D_1$ gauge nodes with edge multiplicity $\fn$. This is precisely in agreement with the description presented below \cite[(6.18)]{Carta:2021whq}.

This can, of course, be generalized to $G$ being $E_6$, $E_7$ and $E_8$. An important point is to choose a basis for the positive roots such that their entries are integer-valued.\footnote{The counting of matter hypermultiplets can also be performed within a M-theoretic setup and is currently under investigation. This method is largely independent from and complementary to ours and leads to the same answer. We thank Mario De Marco, Andrea Sangiovanni and Roberto Valandro for sharing their preliminary results with us. See \cite{Collinucci:2021ofd, DeMarco:2021try} for a detailed discussion about this approach.}

We demonstrate our choice for this in \cref{sec:mirE6mass6,sec:mirrE7mass7,sec:mirrE8mass8}.

\section{$(A_n,E_m)$ theories}
\label{sec:MirrorAE}
In this section, we discuss the $(A_n, E_{6,7,8})$ theories as well as their mirrors in detail.  We find that it is convenient to arrange the discussion according to the number of mass parameters in descending order.

\subsection{3d mirror theories for $(A_n,E_6)$}
\label{sec:mirrE6}
Each of the $(A_n,E_6)$ theories either has $6$, $2$ or $0$ mass parameters.

\subsubsection{6 mass parameters}
\label{sec:mirE6mass6}
This is actually the class of theories discussed earlier in Section \ref{sec:genprophG}, with $G=E_6$, $\rank G = 6$ and $h^\vee_G=12$. They can be parametrized as $(A_{12\fn-1}, E_6)$, with $\fn \geq 1$. One may write the $36$ positive roots $\vec r^{(j)}$ (with $j=1,\ldots 36$) of $E_6$ in terms of the simple roots\footnote{In the standard orthogonal basis $\{\vec e_1, \cdots, \vec e_6\}$, the simple roots of $E_6$ can be written as (see \eg~ \cite[Page 333]{Fulton_2004} and \cite[Appendix A.1]{Keller:2011ek})
\begin{equation*}
\begin{split}
 \vec\alpha_1&= \frac{1}{2} \left(\vec e_1 - \vec e_2 - \vec e_3 - \vec e_4 -\vec e_5 + \sqrt3 \vec e_6\right)~, \\
 \vec\alpha_i &= \vec e_i - \vec e_{i-1}  \ (i=2,\ldots,5), \quad\vec \alpha_6 = \vec e_1+ \vec e_2~.  \label{E6simpleroots}
\end{split}
\end{equation*}
} $\vec \alpha_i$ (with $i=1,\ldots 6$) as
\bes{
\vec r^{(j)} = \sum_{i=1}^6 c^{(j)}_i \vec \alpha_i~, \quad \text{with} ~j=1, \ldots 36
}
with $c^{(j)}_i$ {\it non-negative integers}. The coefficients $c^{(j)}_i$ have an interpretation as the charges of the $j$-th hypermultiplet (in each copy of the $\fn$ copies of such hypermultiplets) under the $i$-th $\U(1)$ gauge group.  These coefficients are, of course, {\it not} unique, but any two choices of the positive roots are related by the Weyl symmetry. The coefficients $c^{(j)}_i$ can be computed, for example, by using the command {\tt AlphaBasis[PositiveRoots[E6]]} of the Mathematica application {\tt LieART 2.0} \cite{Feger:2019tvk}; the result is
\begin{alignat}{4}
(1, 2, 3, 2, 1, 2),\, & (1, 2, 3, 2, 1, 1),\, & (1, 2, 2, 2, 1, 1),\, & (1, 2, 2, 1, 1, 1),\nonumber\\
(1, 1, 2, 2, 1, 1),\, & (0, 1, 2, 2, 1, 1),\, & (1, 2, 2, 1, 0, 1),\, & (1, 1, 2, 1, 1, 1),\nonumber\\
(0, 1, 2, 1, 1, 1),\, & (1, 1, 2, 1, 0, 1),\, & (1, 1, 1, 1, 1, 1),\, & (0, 1, 2, 1, 0, 1),\nonumber\\
(0, 1, 1, 1, 1, 1),\, & (1, 1, 1, 1, 0, 1),\, & (1, 1, 1, 1, 1, 0),\, & (0, 1, 1, 1, 0, 1),\nonumber\\
(0, 1, 1, 1, 1, 0),\, & (0, 0, 1, 1, 1, 1),\, & (1, 1, 1, 0, 0, 1),\, & (1, 1, 1, 1, 0, 0),\\
(0, 1, 1, 0, 0, 1),\, & (0, 1, 1, 1, 0, 0),\, & (0, 0, 1, 1, 0, 1),\, & (0, 0, 1, 1, 1, 0),\nonumber\\
(1, 1, 1, 0, 0, 0),\, & (0, 1, 1, 0, 0, 0),\, & (0, 0, 1, 0, 0, 1),\, & (0, 0, 1, 1, 0, 0),\nonumber\\
(0, 0, 0, 1, 1, 0),\, & (1, 1, 0, 0, 0, 0),\, & (0, 1, 0, 0, 0, 0),\, & (0, 0, 1, 0, 0, 0),\nonumber\\
(0, 0, 0, 0, 0, 1),\, & (0, 0, 0, 1, 0, 0),\, & (0, 0, 0, 0, 1, 0),\, & (1, 0, 0, 0, 0, 0)\coma\nonumber
\end{alignat}
where the entries of the $j$-th vector denote the coefficient $c^{(j)}_{i=1,\ldots,6}$.

\subsubsection{2 mass parameters}
\label{sec:mirrE6mass2}

The $(A_n, E_6)$ theories with 2 mass parameters can be parameterized as $(A_{3\fn-1},E_6)$, for $\fn \geq 1$ and 
\begin{equation}
    3\fn \, \, \mod \, 12 \neq 0\fstop
\end{equation}
We observe that each theory in this class has rank $9\fn-4$, and for $\fn$ of the form $\fn = 4\fm -2$, with $\fm \in \BZ_{\geq 1}$, the theory $(A_{12\fm-7}, E_6)$ has a $\BZ_2$ $1$-form symmetry; otherwise, the $1$-form symmetry is trivial. 

We propose that the mirror theory is  
\bes{ \label{freehyp2masses}
\text{$3\fn-2$ free hypermultiplets}
}
together with an interacting part described by one of the following equivalent descriptions:\footnote{We remark that the interacting part of the $3$d mirror of $(A_{3\mathfrak{n}-1}, E_6)$ for $3\fn \, \, \mod \, 12 \neq 0$ is identical to the $3$d mirror of $(A_2, A_{6\fn -1})$. This in particular implies that the Higgs branch of these two theories is identical. It is interesting to ask whether the two $4$d theories are somehow related.}
\bes{ \label{quiv2massesE6}
&\begin{tikzpicture}[baseline]
\node[draw=none] at (0,0) {$D_1$}; 
\draw[solid, blue,thick] (0.3,0)--(1.5,0) node[above,midway] {$2\fn$};
\draw[solid,snake it]
(2.1,0)--(3.1,0);
\node[draw=none] at (1.8,0) {$D_1$};
\node[draw=none] at (3.6,0) {$[2\fn]_2$};
\end{tikzpicture} \quad /\BZ_2 \\
\qquad \longleftrightarrow \qquad
&\begin{tikzpicture}[baseline]
\node[draw=none] at (0,0) {$\U(1)$}; 
\draw[solid, black,thick] (0.5,0)--(1.3,0) node[above,midway] {$2\fn$};
\draw[solid]
(2.3,0)--(3.1,0);
\draw[solid]
(-1.4,0)--(-0.6,0);
\node[draw=none] at (1.8,0) {$\U(1)$};
\node[draw=none] at (3.6,0) {$[2\fn]$};
\node[draw=none] at (-1.8,0) {$[2\fn]$};
\end{tikzpicture} \\
\qquad \longleftrightarrow \qquad
&\begin{tikzpicture}[baseline]
\node[draw=none] (node1)at (0,0) {$\U(1)$}; 
\node[draw=none] (node2) at (2,0) {$\U(1)$};
\node[draw=none] (node3) at (1,1.5) {$\U(1)$};
\draw[solid, black,thick] (node1)--(node2) node[below,midway] {$2\fn$};
\draw[solid, black,thick] (node2)--(node3) node[above right,midway] {$2\fn$};
\draw[solid, black,thick] (node3)--(node1) node[above left,midway] {$2\fn$};
\end{tikzpicture} \quad /\U(1)
}

Let us now discuss the non-Higgsable sector of this class of theory, as well as the number of free hypermultiplets in the mirror theory. First, the fractional part of $24(c-a)$ contains information about the non-Higgsable sector of the theory.  Since the number of mass parameters is equal to the dimension of the Coulomb branch of the mirror theory which is also equal to the dimension of the Higgs branch of the 4d theory, it follows that
\bes{ \label{24cmaNH}
& \left(\text{$24(c-a)$ of the AD theory} \right) - \left(\text{\# mass parameters of the AD theory}\right) \\
&= \text{$24(c-a)$ of the non-Higgsable sector}~.
}
Moreover, as we discussed earlier, the $1$-form symmetry of the non-Higgsable sector has to be equal to that of the AD theory in question. Finally, the rank of the non-Higgsable sector must be in agreement with the number of free hypermultiplets. As we will see, these three conditions provide a strong constraint on the non-Higgsable sector of a given AD theory.  Let us first examine the theories with a $\BZ_2$ $1$-form symmetry and two mass parameters. We take, for example, $(A_5, E_6)$, $(A_{17}, E_6)$ and $(A_{29}, E_6)$; their quantities \eref{24cmaNH} are, respectively, $0$, $4/5$ and $8/7$.  Non-Higgsable theories with a $\BZ_2$ $1$-form symmetry and these values are $(A_2, D_4)$, $(A_8, D_4)$, and $(A_{14},D_4)$.  The first two can be found in Table \ref{tab:AnDmNHC}, whereas the fact that the last theory is non-Higgsable is described below \cite[(5.4)]{Carta:2021whq}.  The rank of these non-Higgsable theories are $4$, $16$ and $28$; in agreement with \eref{freehyp2masses} for $\fn = 2, \, 6, \, 10$. This observation leads us to the following conclusion:
\begin{quote}
For the $(A_{12\fm-7},E_6)$ theory ($\fm \geq 1$), which has two mass parameters and a $\BZ_2$ $1$-form symmetry, the non-Higgsable theory is $(A_{6\fm-4}, D_4)$.  The latter has a $\BZ_2$ $1$-form symmetry and rank $12\fm-8$.
\end{quote}
Let us turn to the theories with a trivial $1$-form symmetry and two mass parameters. Examples of such theories are $(A_2, E_6)$, $(A_8, E_6)$, $(A_{14}, E_6)$ and $(A_{20}, E_6)$.  Their quantities \eref{24cmaNH} are, respectively, $1/5$, $5/7$, $1$ and $13/11$. For the first two theories, namely $(A_2, E_6)$, $(A_8, E_6)$, the non-Higgsable SCFTs can be easily identified as $(A_1, A_2)$ and $(A_2, E_7)$; their ranks are $1$ and $7$, in agreement with \eref{freehyp2masses}.  For the third and fourth theories, we conjecture that the non-Higgsable sector is a product of non-Higgsable SCFTs with $0<24(c-a)<1$. We, however, cannot identify them reliably using our current technique. We leave this issue for the future investigation.

Let us now discuss some special cases in detail.  This also serves as a non-trivial test of the above proposal.

\subsubsection*{The case of $\fn=1$: The $(A_2,E_6)$ theory}
It can be checked that the Seiberg-Witten curves and differential, Coulomb branch spectra and central charges of the $(A_2,E_6)$ and $(A_3, D_4)$ are equal.  The mirror for the latter is given by \cite[Eqs. (5.16), (5.17) and (5.18)]{Carta:2021dyx} with $\fn=1$, and by \cite[(6.37)]{Carta:2021whq} with $\fm=2, \fN=2$.  This is in agreement with the above proposal.

\subsubsection*{The case of $\fn=2$: The $(A_5,E_6)$ theory}
It was proposed in \cite[left diagram of (3.20)]{Closset:2020afy} that, upon dimensional reduction to 3d, we obtain the following 3d $\CN=4$ quiver gauge theory
\bes{ \label{redA5A6}
\begin{tikzpicture}[baseline,font=\footnotesize]
\node[draw=none] (U1) at (-4,0) {$\U(1)$};
\node[draw=none] (U2) at (-2,0) {$\U(2)$};
\node[draw=none] (SU4) at (0,0) {$\SU(4)$};
\node[draw=none] (U3) at (2,0) {$\U(3)$};
\node[draw=none] (U2p) at (4,0) {$\U(2)$};
\node[draw=none] (SU2) at (6,0) {$\SU(2)$};
\node[draw=none] (U1p) at (8,0) {$\U(1)$};
\node[draw=none] (SU2p) at (0,1.5) {$\SU(2)$};
\draw (U1)--(U2)--(SU4)--(U3)--(U2p)--(SU2)--(U1p);
\draw (SU4)--(SU2p);
\end{tikzpicture}
}
with the possibility of $\BZ_2$ discrete gauging.  The choice of such discrete gauging reflects the presence of the $\BZ_2$ $1$-form global symmetry of the 4d $(A_5, E_6)$ theory that can be either gauged or left ungauged (cf. \cite[Section 5.3]{Carta:2021whq}). We show that the aforementioned theory, namely \eref{freehyp2masses} with \eref{quiv2massesE6}, is indeed a mirror theory for \eref{redA5A6} {\it with} the $\BZ_2$ discrete gauging.  A way to do so is to compute the Coulomb branch Hilbert series of \eref{redA5A6}, which is outlined in Appendix \ref{app:3dA5E6CBHS}.  We find that the contribution from the integral-valued magnetic fluxes is
\bes{
1 + 58 t + 256 t^{3/2} + \ldots~,
}
whereas, if there is $\BZ_2$ discrete gauging, one also needs to take into account the contribution from the half-odd-integral-valued magnetic fluxes:
\bes{
8 t^{1/2} + 24 t + 360 t^{3/2} + \ldots~.
}
Summing up these two contributions, we obtain
\bes{ \label{CBHSredA5E6}
& 1 + 8 t^{1/2} + 82 t + 616 t^{3/2} + \ldots \\
& = \left(1-t^{1/2}\right)^{-8} \left( 1 + 46 t + 128 t^{3/2} + \ldots \right)~.
}
The factor $\left(1-t^{1/2}\right)^{-8}$ indicates the presence of $8$ free half-hypermultiplets, \ie~ 4 free hypermultiplets, which arise from the reduction of the non-Higgsable theory $(A_2, D_4)$. Furthermore, the term $46 t$ indicates that the Coulomb branch symmetry of the interacting part of \eref{redA5A6} is of dimension $46$; this is perfectly in agreement with the Higgs branch symmetry of $\eref{quiv2massesE6}_{\fn=2}$, namely $\SU(4)^3 \times \U(1)$, which has dimension $46$. Moreover, we check that the Higgs branch Hilbert series of $\eref{quiv2massesE6}_{\fn=2}$ is in agreement with the quantity in the brackets in the second line of \eref{CBHSredA5E6}.

\subsubsection*{Maruyoshi-Song flow and the mirror of $D_{3\fn+12}(E_6)$}

As we have already explained, the theories with two mass parameters can be obtained by activating a nilpotent VEV for the $E_6$ moment map of $D_{3\fn+12}(E_6)$. The 3d mirrors of the latter theories are then simply given by $T(E_6)$ coupled to the quiver \eqref{quiv2massesE6}, where coupling means that the two abelian gauge groups should be identified with a subgroup of the Cartan of $E_6$. In order to specify the 3d mirrors of the parent $D_{3\fn+12}(E_6)$ models, we should therefore identify the embedding. This can be done by considering the MS flow, as we will now explain. 

The important fact for us is that $D_{3\fn+12}(E_6)$ can flow both to $(A_{3\fn-1},E_6)$ and to $(A_{3\fn+11},E_6)$ upon nilpotent higgsing and MS flow, respectively. In the first case at the level of the 3d mirror we are simply removing the $T(E_6)$ sector, whereas in the second we are trading it for a chiral multiplet in the adjoint of $E_6$. The components of this chiral are organized in hypermultiplets in one-to-one correspondence with the positive roots of $E_6$ and their charge under the two abelian groups will clearly depend on how they are embedded inside $E_6$. 

Since the difference between nilpotent higgsing and MS flow is a shift of $\fn$ by 4, from \eqref{freehyp2masses} we conclude that the subgroup of $E_6$ commuting with the $\U(1)^2$ gauge group has twelve positive roots. We also know that it has rank $4$ (neglecting the gauge group itself) and therefore we are looking for a group of rank 4 and dimension 28. These are the data of $\SO(8)$ and therefore we conclude that we have to gauge the $\U(1)^2$ subgroup which commutes with $\SO(8)$ inside $E_6$. We can further test this proposal by looking at the decomposition of the adjoint under 
\begin{equation*}
    E_6\supset \U(1)_1\times \SO(10)\supset \U(1)_1\times \U(1)_2\times \SO(8)\,.
\end{equation*} 
Apart from the adjoint of $\SO(8)$, we get a vector of $\SO(8)$ which is charged under $\U(1)_2$ only and two spinors in ${\bf 8_s}$ and ${\bf 8_c}$ which are charged under both $\U(1)$'s, with charge $(1,1)$ and $(1,-1)$ respectively. These data are clearly compatible with a shift of $\fn$ by 4 in the first quiver in \eqref{quiv2massesE6}, giving further support to our claim.

\subsubsection{0 mass parameter}
\label{sec:mirrE6mass0}

The $(A_n, E_6)$ theory with $n$ of the form $n=2 \fn$ (with $\fn\geq 1$) or $n=2\fn-1$ (with $\fn\geq 1$, with $2\fn \, \mod \, 3 \neq 0$ and $2\fn \, \mod \, 4 \neq 0$) has zero mass parameter and a trivial $1$-form symmetry. On the other hand, for $n$ of the form $n=4\fm-1$, with $\fm \geq 1$ and $4\fm \, \mod \, 12 \neq 0$, the theory has zero mass parameter and a $\BZ_3$ $1$-form symmetry.  In any case, the theory is non-Higgsable. The mirror theory is therefore a collection of free hypermultiplets, whose number is equal to the rank of the corresponding 4d theory.  In the case of the theories with a $\BZ_3$ $1$-form symmetry, there is a choice of applying a $\BZ_3$ discrete gauging to such free hypermultiplets in the mirror theory (cf. \cite[Section 5.3]{Carta:2021whq}).

Let us demonstrate the above statement in a non-trivial example of $(A_3, E_6)$. The 3d $\CN=4$ gauge theory description upon dimensional reduction to 3d is given by \cite[(3.13)]{Closset:2020afy}:
\bes{ \label{redA3E6}
\begin{tikzpicture}[baseline,font=\footnotesize]
\node[draw=none] (U1) at (-4,0) {$\U(1)$};
\node[draw=none] (U2) at (-2,0) {$\U(2)$};
\node[draw=none] (SU3) at (0,0) {$\substack{\U(3) \\ \text{or} \\\SU(3)}$};
\node[draw=none] (U2p) at (2,0) {$\U(2)$};
\node[draw=none] (U1p) at (4,0) {$\U(1)$};
\node[draw=none] (U1pp) at (0,1.5) {$\U(1)$};
\draw (U1)--(U2)--(SU3)--(U2p)--(U1p);
\draw (SU3)--(U1pp);
\end{tikzpicture}
}
There are two choices for the central gauge node, namely $\U(3)$ or $\SU(3)$.  

Let us first analyze the case in which it is $\U(3)$.  The above quiver has an overall $\U(1)$ that decouples.  After modding this out, say from the central $\U(3)$ gauge node, it becomes $\U(3)/\U(1) \cong \SU(3)/\BZ_3$. From this perspective, the corresponding quiver theory has a $\BZ_3$ discrete $0$-form global symmetry. According to \cite{Benini:2010uu}, the theory \eref{redA3E6} is then the mirror theory of the 4d class $\CS$ theory associated with $A_2$ sphere with two maximal and one minimal punctures. The latter is indeed a theory of $9$ free hypermultiplets. Hence, with this choice of the central gauge node, we conclude that the mirror theory of $(A_3, E_6)$ is the theory of $9$ free hypermultiplets.  Alternatively, we can apply repeatedly the following duality \cite{Gaiotto:2008ak} to quiver \eref{redA3E6} with the central node being $U(3)$:
\bes{ \label{dualityGW}
&\text{$\U(N)$ gauge theory with $2N-1$ flavors}  \\
\quad \longleftrightarrow \quad &\text{$\U(N-1)$ gauge theory with $2N-1$ flavors} \\
&\text{+ one twisted hypermultiplet}~. 
}
As a result, we obtain a theory of $9$ twisted hypermultiplets, whose mirror theory is a theory of $9$ free hypermultiplets. 

Let us now turn to the other choice of the central node, which is $\SU(3)$. This can be obtained simply by gauging the $\BZ_3$ $0$-form global symmetry of the aforementioned theory.  As a result, the corresponding mirror theory of $(A_3, E_6)$ is a $\BZ_3$ discrete gauging of the theory of $9$ free hypermultiplets. As we discussed earlier, These two choices reflect the presence of the $\BZ_3$ $1$-form symmetry of the $(A_3, E_6)$ theory.

\subsection{3d mirror theories for $(A_n,E_7)$}
\label{sec:mirrE7}

\subsubsection{7 mass parameters}
\label{sec:mirrE7mass7}

This is actually the class of theories discussed earlier in Section \ref{sec:genprophG}, with $G=E_7$, $\rank G = 7$ and $h^\vee_G=18$. Each theory in this class can be written as $(A_{18 \fn-1}, E_7)$.  One may write the $63$ positive roots $\vec r^{(j)}$ (with $j=1,\ldots 63$) of $E_7$ in terms of the simple roots\footnote{The simple roots of $E_7$ in the standard orthogonal basis can be found in \cite[Page 333]{Fulton_2004} written as
\begin{equation*}
\begin{split}
     \vec\alpha_1&= \frac{1}{2} \left(\vec e_1 - \vec e_2 - \vec e_3 - \vec e_4 -\vec e_5 -\vec e_6+ \sqrt2 \vec e_7\right)~, \\
 \vec\alpha_i &= \vec e_i - \vec e_{i-1}  \ (i=2,\ldots,6)\coma\vec \alpha_7 = \vec e_1+ \vec e_2~.  \label{E7simpleroots}
\end{split}
 \end{equation*}
} $\vec \alpha_i$ (with $i=1,\ldots 7$) as
\bes{
\vec r^{(j)} = \sum_{i=1}^7 c^{(j)}_i \vec \alpha_i~, \quad \text{with} ~j=1, \ldots 63
}
with $c^{(j)}_i$ {\it non-negative integers}. As before, the coefficients $c^{(j)}_i$ have an interpretation as the charges of the $j$-th hypermultiplet (in each copy of the $\fn$ copies of such hypermultiplets) under the $i$-th $\U(1)$ gauge group.  We emphasize again that these coefficients are {\it not} unique, but any two choices of the positive roots are related by the Weyl symmetry. The coefficients $c^{(j)}_i$ can be computed, for example, by using the command {\tt AlphaBasis[PositiveRoots[E7]]} of the Mathematica application {\tt LieART 2.0} \cite{Feger:2019tvk}; the result is
\begin{alignat}{4} \label{posrootsE7}
(2, 3, 4, 3, 2, 1, 2),\, & (1, 3, 4, 3, 2, 1, 2),\, & (1, 2, 4, 3, 2, 1, 2),\, & (1, 2, 3, 3, 2, 1, 2),\nonumber \\
(1, 2, 3, 2, 2, 1, 2),\, & (1, 2, 3, 3, 2, 1, 1),\, & (1, 2, 3, 2, 1, 1, 2),\, & (1, 2, 3, 2, 2, 1, 1),\nonumber \\
(1, 2, 3, 2, 1, 0, 2),\, & (1, 2, 3, 2, 1, 1, 1),\, & (1, 2, 2, 2, 2, 1, 1),\, & (1, 2, 3, 2, 1, 0, 1),\nonumber \\
(1, 2, 2, 2, 1, 1, 1),\, & (1, 1, 2, 2, 2, 1, 1),\, & (0, 1, 2, 2, 2, 1, 1),\, & (1, 2, 2, 2, 1, 0, 1),\nonumber \\
(1, 2, 2, 1, 1, 1, 1),\, & (1, 1, 2, 2, 1, 1, 1),\, & (0, 1, 2, 2, 1, 1, 1),\, & (1, 2, 2, 1, 1, 0, 1),\nonumber \\
(1, 1, 2, 2, 1, 0, 1),\, & (1, 1, 2, 1, 1, 1, 1),\, & (0, 1, 2, 2, 1, 0, 1),\, & (0, 1, 2, 1, 1, 1, 1),\nonumber \\
(1, 2, 2, 1, 0, 0, 1),\, & (1, 1, 2, 1, 1, 0, 1),\, & (1, 1, 1, 1, 1, 1, 1),\, & (0, 1, 2, 1, 1, 0, 1),\nonumber \\
(0, 1, 1, 1, 1, 1, 1),\, & (1, 1, 2, 1, 0, 0, 1),\, & (1, 1, 1, 1, 1, 0, 1),\, & (1, 1, 1, 1, 1, 1, 0), \\
(0, 1, 2, 1, 0, 0, 1),\, & (0, 1, 1, 1, 1, 0, 1),\, & (0, 1, 1, 1, 1, 1, 0),\, & (0, 0, 1, 1, 1, 1, 1),\nonumber \\
(1, 1, 1, 1, 0, 0, 1),\, & (1, 1, 1, 1, 1, 0, 0),\, & (0, 1, 1, 1, 0, 0, 1),\, & (0, 1, 1, 1, 1, 0, 0),\nonumber \\
(0, 0, 1, 1, 1, 0, 1),\, & (0, 0, 1, 1, 1, 1, 0),\, & (1, 1, 1, 0, 0, 0, 1),\, & (1, 1, 1, 1, 0, 0, 0),\nonumber \\
(0, 1, 1, 0, 0, 0, 1),\, & (0, 1, 1, 1, 0, 0, 0),\, & (0, 0, 1, 1, 0, 0, 1),\, & (0, 0, 1, 1, 1, 0, 0),\nonumber \\
(0, 0, 0, 1, 1, 1, 0),\, & (1, 1, 1, 0, 0, 0, 0),\, & (0, 1, 1, 0, 0, 0, 0),\, & (0, 0, 1, 0, 0, 0, 1),\nonumber \\
(0, 0, 1, 1, 0, 0, 0),\, & (0, 0, 0, 1, 1, 0, 0),\, & (0, 0, 0, 0, 1, 1, 0),\, & (1, 1, 0, 0, 0, 0, 0),\nonumber \\
(0, 1, 0, 0, 0, 0, 0),\, & (0, 0, 1, 0, 0, 0, 0),\, & (0, 0, 0, 0, 0, 0, 1),\, & (0, 0, 0, 1, 0, 0, 0),\nonumber \\
(0, 0, 0, 0, 1, 0, 0),\, & (0, 0, 0, 0, 0, 1, 0),\, & (1, 0, 0, 0, 0, 0, 0),\, & \nonumber
\end{alignat}
where the entries of the $j$-th vector denote the coefficient $c^{(j)}_{i=1,\ldots,7}$.

\subsubsection{1 mass parameter}
\label{sec:mirrE7mass1}

Each theory in this class can be written as $(A_{2\fn-1},E_7)$ with $\fn \geq 1$ and $2\fn \, \mod \, 18 \neq 0$. The rank is $7 \fn -4$.  Moreover, for $\fn = 3\fm$ (with $\fm \geq 1$), the $(A_{6 \fm -1}, E_7)$ theory has a $\BZ_3$ $1$-form symmetry; otherwise, the $1$-form symmetry is trivial. 

We propose that the mirror theory for $(A_{2\fn-1},E_7)$ is described by 
\bes{\label{mirrE7mass1}
&\begin{tikzpicture}[baseline]
\node[draw=none] (A) at (0,0) {$\U(1)$}; 
\node[draw=none] (B) at (2,0) {$[3\fn]$};
\draw[solid, thick] (A)--(B);
\end{tikzpicture}\\
&\text{+ $4\fn-3$ free hypermultiplets.}
}

For the theories with a $\BZ_3$ $1$-form symmetry, \eg~ $(A_5, E_7)$, $(A_{11},E_7)$ and $(A_{23},E_7)$, their quantities \eref{24cmaNH} are respectively $0$, $3/5$ and $9/7$. Using the information about the $1$-form symmetry and $24(c-a)$, we can identify the corresponding non-Higgsable SCFTs as $(A_3, E_6)$ and $(A_7, E_6)$ and $(A_{15}, E_6)$ respectively.  Note that the information about the first two theories can be found in Table \ref{tab:GGpNHC}, whereas $(A_{15}, E_6)$ has zero mass parameters, a $\BZ_3$ $1$-form symmetry, rank $45$, and $24(c-a)=9/7$; these properties fit precisely the requirement to be the non-Higgsable SCFT for $(A_{23},E_7)$. This observation leads us to the following conclusion:
\begin{quote}
For the $(A_{6\fm-1},E_7)$ theory ($\fm \geq 1$ and $6\fm \, \mod \, 18 \neq 0$), which has one mass parameter and a $\BZ_3$ $1$-form symmetry, the non-Higgsable theory is $(A_{4\fm-1}, E_6)$.  The latter has a $\BZ_3$ $1$-form symmetry and rank $12\fm-3$.
\end{quote}
For the theories with a trivial $1$-form symmetry, it is more difficult to identify the non-Higgsable sector. As an example, for $(A_1, E_7)$ and $(A_3, E_7)$, their quantities \eref{24cmaNH} are respectively $1/5$ and $5/11$; these lead us to identify the non-Higgsable SCFTs as $(A_1,A_2)$ and $(A_2,D_5)$, respectively; the latter two theories have rank $1$ and $5$ respectively and no $1$-form symmetry, and so that they are consistent with our proposal.

\subsubsection*{Maruyoshi-Song flow and the mirror of $D_{2\fn+18}(E_7)$} 

As in Section \ref{sec:mirrE6mass2} we can use the MS flow to understand how the $\U(1)$ gauge group in \eqref{mirrE7mass1} is embedded inside $E_7$ and therefore what the 3d mirror of $D_{2\fn+18}(E_7)$ is. We first notice that the difference between nilpotent higgsing and MS flow amounts to a shift of $\fn$ by 9, and therefore we conclude from \eqref{mirrE7mass1} that the commutant of the gauge group inside $E_7$ has 36 positive roots and has rank 6. These are the data of $E_6$ and therefore we conclude that we have to gauge the $\U(1)$ commuting with an $E_6$ subgroup of $E_7$. Furthermore, we can notice that in the decomposition of the adjoint representation, we get the adjoint of $E_6$ (which contributes to the free sector only) and a ${\bf 27}$ charged under the gauged $\U(1)$. This again is compatible with a shift of $\fn$ by 9 in \eqref{mirrE7mass1}.

\subsubsection{0 mass parameter}
\label{sec:mirrE7mass0}

The $(A_n, E_7)$ theory with $n$ of the form $n=2 \fn$ (with $\fn\geq 1$)  has zero mass parameter and a trivial $1$-form symmetry except for $n = 18\fn -10$, for which the theory has zero mass parameter and a $\BZ_2^3$ $1$-form symmetry. In each of these cases, the theory is non-Higgsable. Once again, the mirror theory is a collection of free hypermultiplets, whose number is equal to the rank of the corresponding 4d theory. In the case of the theories with a $\BZ_2^3$ $1$-form symmetry, there is a choice of applying a discrete gauging to such free hypermultiplets in the mirror theory (cf. \cite[Section 5.3]{Carta:2021whq}). 

In the following, we can demonstrate using the method described in Section \ref{sec:1formonCM} that the $1$-form symmetry for $(A_8,E_7)$ is $\BZ_2^3$.  According to \cite{Closset:2020afy}, the $(A_8,E_7)$ is admits the following weakly gauging description:
\bes{ \label{A8E7}
\begin{tikzpicture}[baseline,font=\footnotesize]
\node[draw=none] (U2) at (-2,0) {$D_3^4(\SO(8))$};
\node[draw=none] (SU3) at (0,0) {$\Spin(8)$};
\node[draw=none] (U2p) at (2,0) {$D_9(\SO(8))$};
\node[draw=none] (U1pp) at (0,1.5) {$\SU(2)$};
\draw (U2)--(SU3)--(U2p);
\draw (SU3)--(U1pp);
\end{tikzpicture}
}
Using \cite{Hosseini:2021ged}, we see that $D_9(SO(8))$ has the same 1-form symmetry as $(A_8, D_4)$, namely $\BZ_2$.  On the other hand, the $D_3^4(\SO(8))$ has the same 1-form symmetry as the $D^{(4)}_4[3]$ theory, but this turns out to be trivial \cite{Hosseini:2021ged}. The center of $SU(2)$ is $\BZ_2$, contributing a factor of $\BZ_2$ 1-form symmetry. One $\BZ_2$ factor of the $\BZ_2 \times \BZ_2$ center symmetry of $\Spin(8)$ is screened by the half-hypermultiplet in the $[\mathbf{2};\mathbf{8}_v]$ representation of $\SU(2) \times \Spin(8)$.  The Higgs branch generators of $D_3^4(\SO(8))$ and $D_9(\SO(8))$ do not contain a spinor representation of $\Spin(8)$ (for the former cf. \cite[Section 7.1]{Carta:2021whq}) and for the latter see Section \ref{sec:1formonCM}).  Therefore, we have one $\BZ_2$ factor center symmetry from the $\Spin(8)$ that is left unscreened. In summary, we have the $\BZ_2^3$ 1-form symmetry, as claimed.

\subsection{3d mirror theories for $(A_n,E_8)$}
\label{sec:mirrE8}

\subsubsection{8 mass parameters}
\label{sec:mirrE8mass8}

This is actually the class of theories discussed earlier in Section \ref{sec:genprophG}, with $G=E_8$, $\rank G = 8$ and $h^\vee_G=30$. Each theory in this class can be written as $(A_{30\fn-1} ,E_8)$, with $\fn\geq 1$. One may write the $120$ positive roots $\vec r^{(j)}$ (with $j=1,\ldots 120$) of $E_8$ in terms of the simple roots\footnote{The simple roots of $E_8$ in the standard orthogonal basis can be found in \cite[Page 333]{Fulton_2004} as
\begin{equation*}
\begin{split}
     \vec\alpha_1&= \frac{1}{2} \left(\vec e_1 - \vec e_2 - \vec e_3 - \vec e_4 -\vec e_5 -\vec e_6-\vec e_7+  \vec e_8\right)~, \\
 \vec\alpha_i &= \vec e_i - \vec e_{i-1}  \ (i=2,\ldots,7)\coma\vec \alpha_8 = \vec e_1+ \vec e_2~.  \label{E8simpleroots}
\end{split}
 \end{equation*}
} $\vec \alpha_i$ (with $i=1,\ldots 8$) as
\bes{
\vec r^{(j)} = \sum_{i=1}^8 c^{(j)}_i \vec \alpha_i~, \quad \text{with} ~j=1, \ldots, 120.
}
with $c^{(j)}_i \in \BZ_{\geq 0}$.
As before, the coefficients $c^{(j)}_i$ can be computed by using the command {\tt AlphaBasis[PositiveRoots[E8]]} of the Mathematica application {\tt LieART 2.0} \cite{Feger:2019tvk}; the result is
\begin{alignat}{4} \label{E8posroots}
(2, 4, 6, 5, 4, 3, 2, 3),\, &(2, 4, 6, 5, 4, 3, 1, 3),\, &(2, 4, 6, 5, 4, 2, 1, 3),\, &(2, 4, 6, 5, 3, 2, 1, 3), \nonumber\\
(2, 4, 6, 4, 3, 2, 1, 3),\, &(2, 4, 5, 4, 3, 2, 1, 3),\, &(2, 4, 5, 4, 3, 2, 1, 2),\, &(2, 3, 5, 4, 3, 2, 1, 3), \nonumber\\
(1, 3, 5, 4, 3, 2, 1, 3),\, &(2, 3, 5, 4, 3, 2, 1, 2),\, &(1, 3, 5, 4, 3, 2, 1, 2),\, &(2, 3, 4, 4, 3, 2, 1, 2), \nonumber\\
(1, 3, 4, 4, 3, 2, 1, 2),\, &(2, 3, 4, 3, 3, 2, 1, 2),\, &(1, 3, 4, 3, 3, 2, 1, 2),\, &(1, 2, 4, 4, 3, 2, 1, 2), \nonumber\\
(2, 3, 4, 3, 2, 2, 1, 2),\, &(1, 3, 4, 3, 2, 2, 1, 2),\, &(1, 2, 4, 3, 3, 2, 1, 2),\, &(2, 3, 4, 3, 2, 1, 1, 2), \nonumber\\
(1, 3, 4, 3, 2, 1, 1, 2),\, &(1, 2, 4, 3, 2, 2, 1, 2),\, &(1, 2, 3, 3, 3, 2, 1, 2),\, &(2, 3, 4, 3, 2, 1, 0, 2), \nonumber\\
(1, 3, 4, 3, 2, 1, 0, 2),\, &(1, 2, 4, 3, 2, 1, 1, 2),\, &(1, 2, 3, 3, 2, 2, 1, 2),\, &(1, 2, 3, 3, 3, 2, 1, 1), \nonumber\\
(1, 2, 4, 3, 2, 1, 0, 2),\, &(1, 2, 3, 3, 2, 1, 1, 2),\, &(1, 2, 3, 2, 2, 2, 1, 2),\, &(1, 2, 3, 3, 2, 2, 1, 1), \nonumber\\
(1, 2, 3, 3, 2, 1, 0, 2),\, &(1, 2, 3, 2, 2, 1, 1, 2),\, &(1, 2, 3, 3, 2, 1, 1, 1),\, &(1, 2, 3, 2, 2, 2, 1, 1), \nonumber\\
(1, 2, 3, 2, 2, 1, 0, 2),\, &(1, 2, 3, 2, 1, 1, 1, 2),\, &(1, 2, 3, 3, 2, 1, 0, 1),\, &(1, 2, 3, 2, 2, 1, 1, 1), \nonumber\\
(1, 2, 2, 2, 2, 2, 1, 1),\, &(1, 2, 3, 2, 1, 1, 0, 2),\, &(1, 2, 3, 2, 2, 1, 0, 1),\, &(1, 2, 3, 2, 1, 1, 1, 1), \nonumber\\
(1, 2, 2, 2, 2, 1, 1, 1),\, &(1, 1, 2, 2, 2, 2, 1, 1),\, &(0, 1, 2, 2, 2, 2, 1, 1),\, &(1, 2, 3, 2, 1, 0, 0, 2), \nonumber\\
(1, 2, 3, 2, 1, 1, 0, 1),\, &(1, 2, 2, 2, 2, 1, 0, 1),\, &(1, 2, 2, 2, 1, 1, 1, 1),\, &(1, 1, 2, 2, 2, 1, 1, 1), \nonumber\\
(0, 1, 2, 2, 2, 1, 1, 1),\, &(1, 2, 3, 2, 1, 0, 0, 1),\, &(1, 2, 2, 2, 1, 1, 0, 1),\, &(1, 2, 2, 1, 1, 1, 1, 1), \nonumber\\
(1, 1, 2, 2, 2, 1, 0, 1),\, &(1, 1, 2, 2, 1, 1, 1, 1),\, &(0, 1, 2, 2, 2, 1, 0, 1),\, &(0, 1, 2, 2, 1, 1, 1, 1), \nonumber\\
(1, 2, 2, 2, 1, 0, 0, 1),\, &(1, 2, 2, 1, 1, 1, 0, 1),\, &(1, 1, 2, 2, 1, 1, 0, 1),\, &(1, 1, 2, 1, 1, 1, 1, 1),\\
(0, 1, 2, 2, 1, 1, 0, 1),\, &(0, 1, 2, 1, 1, 1, 1, 1),\, &(1, 2, 2, 1, 1, 0, 0, 1),\, &(1, 1, 2, 2, 1, 0, 0, 1), \nonumber\\
(1, 1, 2, 1, 1, 1, 0, 1),\, &(1, 1, 1, 1, 1, 1, 1, 1),\, &(0, 1, 2, 2, 1, 0, 0, 1),\, &(0, 1, 2, 1, 1, 1, 0, 1), \nonumber\\
(0, 1, 1, 1, 1, 1, 1, 1),\, &(1, 2, 2, 1, 0, 0, 0, 1),\, &(1, 1, 2, 1, 1, 0, 0, 1),\, &(1, 1, 1, 1, 1, 1, 0, 1), \nonumber\\
(1, 1, 1, 1, 1, 1, 1, 0),\, &(0, 1, 2, 1, 1, 0, 0, 1),\, &(0, 1, 1, 1, 1, 1, 0, 1),\, &(0, 1, 1, 1, 1, 1, 1, 0), \nonumber\\
(0, 0, 1, 1, 1, 1, 1, 1),\, &(1, 1, 2, 1, 0, 0, 0, 1),\, &(1, 1, 1, 1, 1, 0, 0, 1),\, &(1, 1, 1, 1, 1, 1, 0, 0), \nonumber\\
(0, 1, 2, 1, 0, 0, 0, 1),\, &(0, 1, 1, 1, 1, 0, 0, 1),\, &(0, 1, 1, 1, 1, 1, 0, 0),\, &(0, 0, 1, 1, 1, 1, 0, 1), \nonumber\\
(0, 0, 1, 1, 1, 1, 1, 0),\, &(1, 1, 1, 1, 0, 0, 0, 1),\, &(1, 1, 1, 1, 1, 0, 0, 0),\, &(0, 1, 1, 1, 0, 0, 0, 1), \nonumber\\
(0, 1, 1, 1, 1, 0, 0, 0),\, &(0, 0, 1, 1, 1, 0, 0, 1),\, &(0, 0, 1, 1, 1, 1, 0, 0),\, &(0, 0, 0, 1, 1, 1, 1, 0), \nonumber\\
(1, 1, 1, 0, 0, 0, 0, 1),\, &(1, 1, 1, 1, 0, 0, 0, 0),\, &(0, 1, 1, 0, 0, 0, 0, 1),\, &(0, 1, 1, 1, 0, 0, 0, 0), \nonumber\\
(0, 0, 1, 1, 0, 0, 0, 1),\, &(0, 0, 1, 1, 1, 0, 0, 0),\, &(0, 0, 0, 1, 1, 1, 0, 0),\, &(0, 0, 0, 0, 1, 1, 1, 0), \nonumber\\
(1, 1, 1, 0, 0, 0, 0, 0),\, &(0, 1, 1, 0, 0, 0, 0, 0),\, &(0, 0, 1, 0, 0, 0, 0, 1),\, &(0, 0, 1, 1, 0, 0, 0, 0), \nonumber\\
(0, 0, 0, 1, 1, 0, 0, 0),\, &(0, 0, 0, 0, 1, 1, 0, 0),\, &(0, 0, 0, 0, 0, 1, 1, 0),\, &(1, 1, 0, 0, 0, 0, 0, 0), \nonumber\\
(0, 1, 0, 0, 0, 0, 0, 0),\, &(0, 0, 1, 0, 0, 0, 0, 0),\, &(0, 0, 0, 0, 0, 0, 0, 1),\, &(0, 0, 0, 1, 0, 0, 0, 0), \nonumber\\
(0, 0, 0, 0, 1, 0, 0, 0),\, &(0, 0, 0, 0, 0, 1, 0, 0),\, &(0, 0, 0, 0, 0, 0, 1, 0),\, &(1, 0, 0, 0, 0, 0, 0, 0)\coma \nonumber
\end{alignat}
where the entries of the $j$-th vector denote the coefficient $c^{(j)}_{i=1,\ldots,8}$.  We emphasize again the non-uniqueness of the coefficients $c^{(j)}_{i}$.  Note that this choice of basis leads to hypermultiplets of higher charges, \eg $3$, $4$, $5$ and $6$. One can choose another basis such that the charges have lower values, \eg~ in the basis of the fundamental weights given by the command {\tt OmegaBasis[PositiveRoots[E8]]} of {\tt LieART 2.0}, the charges of the hypermultiplets take values $0$, $\pm 1$ and $\pm 2$.  As is well known, the positive roots and the fundamental weights are related to each other by the Cartan matrix.

\subsubsection{0 mass parameter}
\label{sec:mirrE8mass0}

The $(A_n, E_8)$ theory with $n$ {\it not} of the form $18\fn-1$, with $\fn\geq 1$, has zero mass parameter.  The theories with a non-trivial $1$-form symmetry are as follows.
\bi
\item Those with a $\BZ_5$ $1$-form symmetry are $(A_{6\fn-1}, E_8)$ with $6\fn \, \mod \, 30 \neq 0$.  
\item Those with a $\BZ_3^2$ $1$-form symmetry are  $(A_{10\fn-1}, E_8)$ with $10\fn \, \mod \, 6 \neq 0$.
\item Those with a $\BZ_2^4$ $1$-form symmetry are
$(A_{15\fn-1}, E_8)$ with $15\fn \, \mod \, 6 \neq 0$.
\ei
The theories with zero mass parameter that do not belong to the above classes have a trivial $1$-form symmetry.  Note that every theory with zero mass parameter is non-Higgsable. We conclude that the mirror theory is a collection of free hypermultiplets, whose number is equal to the rank of the corresponding 4d theory. In the case of the theories with a non-trivial $1$-form symmetry, there is an option for applying the discrete gauging of such a $1$-form symmetry to the free hypermultiplets (cf. \cite[Section 5.3]{Carta:2021whq}). Let us check the above statement in the special case of the $(A_5,E_8)$ theory which belongs to the series $(A_{6\fn-1}, E_8)$ described before. We know that this theory is equivalent to a $\SU(5)$ gauge theory coupled to $D_2(\SU(5))$, $D_3(\SU(5))$ and $D_6(\SU(5))$. Using the results derived in \cite{Closset:2020afy} we know that upon dimensional reduction $(A_5,E_8)$ becomes the following quiver: 
\bes{ \label{redA5E8}
\begin{tikzpicture}[baseline,font=\footnotesize]
\node[draw=none] (U1) at (-4,0) {$\U(1)$};
\node[draw=none] (U3) at (-2,0) {$\U(3)$};
\node[draw=none] (SU5) at (0,0) {$\substack{\U(5) \\ \text{or} \\\SU(5)}$};
\node[draw=none] (U4p) at (2,0) {$\U(4)$};
\node[draw=none] (U3p) at (4,0) {$\U(3)$};
\node[draw=none] (U2p) at (6,0) {$\U(2)$};
\node[draw=none] (U1p) at (8,0) {$\U(1)$};
\node[draw=none] (U2pp) at (0,1.5) {$\U(2)$};
\draw (U1)--(U3)--(SU5)--(U4p)--(U3p)--(U2p)--(U1p);
\draw (SU5)--(U2pp);
\end{tikzpicture}
}
Let us consider the case in which the central node is taken to be $\U(5)$. We see that the quiver becomes ugly.  Using repeatedly the duality \eref{dualityGW}, we see that the above quiver theory is equivalent to a collection of 20 twisted hypermultiplets, in agreement with our claim. Another way to see this is to regard \eref{redA5E8} as a mirror theory of the class $\CS$ theory associated with $A_4$ sphere with the punctures $[1^5]$, $[2^2, 1]$ and $[3,2]$.  This can be identified with the theory of 20 free hypermultiplets \cite[Page 30]{Chacaltana:2010ks}.  In this case, there is an overall $\U(1)$ needs to be modded out from quiver \eref{redA5E8}; this results in the central node $\U(5)/\U(1) \cong \SU(5)/\BZ_5$, which leads to a $\BZ_5$ $0$-form global symmetry.  Let us turn to the case in which the central node is $\SU(5)$.  This is simply the gauging of the aforementioned $\mathbb{Z}_5$ $0$-form symmetry. This results in the mirror theory, which is a $\BZ_5$ discrete gauging of 20 free hypermultiplets.

\subsection{A comment on 2-group structures}

Let $F$ denote the $0$-form flavor symmetry group of a $4d$ $\mathcal{N}=2$ theory, and $\Gamma^{(1)}$ denote the $1$-form symmetry group. Let BF denote the classifying space of F, where the latter is seen as a topological space. A necessary condition in order to have $2$-group structure is that the Postnikov class $w_3\in H^3(\text{BF},\Gamma^{(1)})$ is non-vanishing (see \eg \cite{Bhardwaj:2021wif}). In this section we argue that for Argyres-Douglas theories of type $(A_n, E_m)$, the cohomology group $H^3(\text{BF},\Gamma^{(1)})$ is actually trivial, therefore such theories do not enjoy $2$-group symmetries.\footnote{We refer the readers to \cite{Bah:2022xfv, DelZotto:2022joo, Cvetic:2022imb} for recent developments on 2-group structures.}

Let us start by recalling which of those theories admit $1$-form symmetry as well as the number of mass parameters, which corresponds to the rank of the $0$-form flavor symmetry. This information is summarized in Table \ref{tab:1form-masses}. Whenever we have a rank zero $0$-form flavor symmetry, we make the working assumption that such flavor symmetry is trivial. In particular, we assume there are no discrete $0$-form symmetries.

\begin{table}[!htp]
\centering
\begin{tabular}{c|cl|c}
AD theory & \multicolumn{2}{c|}{$1$-form symmetry group $ \Gamma^{(1)}$}& $\#$ mass parameters                                                                                                                                          \\ \hline
$(A_{n-1}, E_6)$                 & \begin{tabular}[c]{@{}c@{}}$0$\\ $\mathbb{Z}_2$\\ $\mathbb{Z}_3$\\ $0$\end{tabular}                                   & \multicolumn{1}{l|}{\begin{tabular}[c]{@{}l@{}}if $12 \mid n$\\ if $6 \mid n$\\ if $4 \mid n$\\ otherwise\end{tabular}} & \multicolumn{1}{l}{\begin{tabular}[c]{@{}l@{}} $6$ \\ $2$\\ $0$\\ varies\end{tabular}}                  \\ \hline
$(A_{n-1}, E_7)$                 & \begin{tabular}[c]{@{}c@{}}$0$\\ $\mathbb{Z}_2^3$\\ $\mathbb{Z}_3$\\ $0$\end{tabular}                                   & \multicolumn{1}{l|}{\begin{tabular}[c]{@{}l@{}}if $18 \mid n$\\ if $9 \mid n$\\ if $6 \mid n$\\ otherwise\end{tabular}} & \multicolumn{1}{l}{\begin{tabular}[c]{@{}l@{}} $7$\\ $0$\\ $1$\\ varies\end{tabular}}                  \\ \hline
$(A_{n-1}, E_8)$                 & \begin{tabular}[c]{@{}c@{}}$0$\\ $\mathbb{Z}_2^4$\\ $\mathbb{Z}_3^2$\\ $\mathbb{Z}_5$\\ $0$ \end{tabular}      & \multicolumn{1}{l|}{\begin{tabular}[c]{@{}l@{}}if $30 \mid n$\\ if $15 \mid n$\\ if $10 \mid n$\\ if $6 \mid n$\\ otherwise\end{tabular}} & \multicolumn{1}{l}{\begin{tabular}[c]{@{}l@{}} $8$\\ $0$\\ $0$\\ $0$\\ varies\end{tabular}} 
\end{tabular}
\caption{$1$-form symmetry groups and number of mass parameters for theories of type $(A_n,E_m)$. For theories where the cases overlap, the highest written condition takes priority.}
\label{tab:1form-masses}
\end{table}

From the data in Table \ref{tab:1form-masses} above, we clearly see that in most cases in which there is present a non-trivial $1$-form symmetry, the flavor is trivial. The only cases in which $1$-form symmetries and $0$-form symmetries coexist, and therefore there is a chance that $H^3(\text{BF}, \Gamma^{(1)})$ is non-trivial, are $(A_{n-1}, E_6)$ with $2$ masses, and $(A_{n-1}, E_7)$ with one mass. We can infer the global form of the $0$-form flavor symmetry group for both those theories by an explicit computation of the refined CB Hilbert series of their $3$d mirror, which has been introduced in \cite{Cremonesi:2013lqa}. Such computation leads to the result that the $0$-form flavor symmetry for $(A_{n-1}, E_6)$  with $2$ masses is $\U(1)^2$, while the one for $(A_{n-1}, E_7)$ with one mass is $\U(1)$.

The question about the existence of $2$-group structure then boils down to the computation of $H^3(\text{BU}(1), \mathbb{Z}_k)$ for $k=2, 3$. It is known in general that the cohomology of the classifying space of $\U(n)$ with integer coefficients, namely, $H^\star(\text{BU}(n),\mathbb{Z})$ is isomorphic to the ring of polynomial in $n$ variables $c_1, \ldots c_n$, where $c_k$ is of degree $2k$; see \eg~ \cite[Page 227]{hatcher2002algebraic}. Note that $H^\star(\text{BU}(n),\mathbb{Z})$ means the direct sum of $H^p(\text{BU}(n),\mathbb{Z})$, and so elements of $H^\star(\text{BU}(n),\mathbb{Z})$ are finite sums $\sum_i \alpha_i$ with $\alpha_i \in H^i(\text{BU}(n),\mathbb{Z})$. Since each $c_k$ is of an even degree, $H^p(\text{BU}(n),\mathbb{Z})$ is trivial for every $n$ and every odd $p$. From an application of the universal coefficient theorem, it follows that $H^3(\text{BU}(1), \mathbb{Z}_k)$ is trivial for all $k\geq 2$. We thus arrive at the conclusion that all $(A_n, E_m)$ Argyres-Douglas theories do not enjoy $2$-group structure.\footnote{We remark that the validity of this conclusion relies on the assumption that no discrete $0$-form symmetries are present in the 4d theory.}

\section{$(D_n,E_m)$ theories}
\label{sec:CommDE}

In this section, we discuss the $(D_n, E_{6,7,8})$ theories. In contrast to the $(A_n, E_{6,7,8})$ theories, the Calabi-Yau hypersurfaces associated with the former contain a crepant divisor. The latter is an exceptional divisor that quantifies how much the hypersurface singularity can be ``smoothened'' by means of the crepant resolution. It has been observed in \cite{
Closset:2020scj, Closset:2020afy, Carta:2021dyx} that, for a given $(G,G')$ theory, the number of crepant divisor is equal to the difference between the Higgs branch dimension and the number of mass parameters. In particular, it was pointed out in \cite{Carta:2021dyx} that the presence of the crepant divisors in the $(D_n, D_m)$ theories lead to the presence of a non-abelian gauge group of the symplectic type in the corresponding mirror theories.

Even though the physical implication of the crepant divisors is well-understood in the context of the $(D_n, D_m)$ theories in our previous work \cite{Carta:2021dyx}, we find that their presence in the $(A_n, E_{6,7,8})$ leads to a much higher level of complication in the Higgs branch structures and in finding the correct mirror theories.  For example, it was pointed out in \cite{Gaiotto:2012uq} that the $(D_4, E_6)$ theory has a ``bad'' class $\CS$ description in the sense of \cite{Gaiotto:2012uq} and that the theory can flow to either to a theory of a collection of free hypermultiplets, or an interacting rank-one $E_6$ theory with a free hypermultiplet.  Such complications prevent us to find general results for $(D_n, E_{6,7,8})$ theories.  In what follows, we only discuss the $(D_{12 \fn-5}, E_6)$ theories, with $\fn \geq 1$, in full generality, and present only partial results for the others.  We hope to revisit the latter in future work.  

Finally, we remark also that, similarly to the $(D_n, D_m)$ theories \cite{Carta:2021dyx}, the $(D_n,E_m)$ theories do not admit a known class $\mathcal{S}$ description. Moreover, one cannot make use of the Maruyoshi-Song flow and Flip-Flip duality to justify the proposed mirror theories.  Thus, for the $(D_{12 \fn-5}, E_6)$ theories, we manage to only provide the test by mass deformations and by gauging the baryonic symmetries for $\fn=1$. We also hope to provide further checks for the proposed mirrors for these theories in the future.

\subsection{3d mirror theories for $(D_n, E_6)$ theories}
Using the method of computing the number of crepant divisor reviewed in \cite[Section 3.2]{Carta:2021dyx},\footnote{The algorithm that computes the number of crepant divisors is described in \cite{Caibar:1999aaa,Caibar2003:aaa}.} we find that the only $(D_n, E_6)$ theory with no crepant divisors is $(D_3,E_6)\cong (A_3, E_6)$, otherwise for $n\geq 4$, there is $1$ crepant divisor.  

\subsubsection{4 mass parameters}
The theories with 4 mass parameters can be written as $(D_{12\fn-5},E_6)$ with $\fn \in \BZ_{\geq 1}$. Each of these theories has $24(c-a)=5$ and rank $36\fn-17$.  

We propose that the mirror theory is described by
\bes{ \label{mirrD12m5E6}
\begin{tikzpicture}[baseline,font=\footnotesize]
\node[draw=none] (F) at (-2,0) {$[D_4]$};
\node[draw=none] (C) at (0,0) {$C_2$};
\node[draw=none] (U11) at (2,0) {$\U(1)$};
\node[draw=none] (U12) at (4,1.5) {$\U(1)$};
\node[draw=none] (U13) at (4,-1.5) {$\U(1)$};
\draw (F)--(C)--(U11);
\draw[black, thick] (U11) to node[above left]{$12\fn-8$} (U12);
\draw[black, thick] (U11) to node[below left]{$12\fn-8$} (U13);
\draw[black, thick] (U12) to node[right]{$12\fn-8$} (U13);
\end{tikzpicture}
}
where each thick black edge has multiplicity $12 \fn-8$ and there is no free hypermultiplet in the mirror theory.  We emphasize that the $C_2$ node has 5 flavors transformed under the fundamental representation, and so it is balanced; as a result, we have one $\U(1)$ global symmetry emergent in the IR. Together with the three $\U(1)$ gauge nodes, we have in total $4$ $\U(1)$ topological symmetries, in correspondence with the 4 mass parameters of the 4d theory.  The mirror theory has 5 dimensional Coulomb branch, in agreement with $24(c-a)$ of the 4d theory, and $36\fn-17$ dimensional Higgs branch, in agreement with the rank of the 4d theory.  In the following, we provide two highly non-trivial test of the proposed mirror theory for $\fn=1$.

\subsubsection*{Test by mass deformations}

Let us discuss further non-trivial tests of $\eref{mirrD12m5E6}_{\fn=1}$.  First, we exploit the fact that $(D_7, E_6)$ can be mass deformed to either $(D_7, D_4)$ or $(A_5, E_6)$, which can be seen as follows: The defining equation of the $(D_7,E_6)$ singularity is 
\be x^6+xy^2+z^3+w^4=0.\ee 
and the scaling dimension of the various coordinates is 
$$D(x)=1;\; D(y)=\frac{5}{2};\; D(z)=2;\; D(w)=\frac{3}{2}.$$ 
We therefore conclude that the following two deformations
\be x^6+xy^2{\color{blue} +my^2}+z^3+w^4{\color{blue} +m'zw^2}=0\ee 
both correspond to mass deformations of the theory, since the parameters $m$ and $m'$ have dimension one. The former induces the flow to $(A_5,E_6)$ and the latter to $(D_7,D_4)$. Both theories have two mass parameters, and therefore we expect to recover their 3d mirrors from \eqref{mirrD12m5E6} (with $\fn=1$) upon turning on two independent FI parameters. Since we already know the 3d mirror of both theories, we are in the position to check this claim explicitly, and we regard the fact that it works as a highly nontrivial test of our proposal. 

The flow to $(D_7,D_4)$ is easy to analyze, since we clearly see that turning on FI parameters $\lambda_1$, $\lambda_2$ and $-\lambda_1-\lambda_2$ at the three abelian nodes in \ref{mirrD12m5E6}, the equations of motion are satisfied by activating a VEV for two $\U(1)\times \U(1)$ bifundamentals and this higgses the theory to \bes{
&[D_4]-C_2- \U(1) \\
& \text{with $10$ free hypermultiplets.}
} 
This is indeed the mirror theory for $(D_7, D_4)$ plus 6 free hypermultiplets; see \cite[(5.11), (5.12) with $n=4, m=7$]{Carta:2021dyx}. The extra hypermultiplets account for the difference in rank between the two theories\footnote{We remark that mass deformations preserve the rank of the theory, so the statement really is that upon a suitably-chosen mass deformation $(D_7,E_6)$ flows to $(D_7,D_4)$ plus 6 free vector multiplets.} (19 for $(D_7,E_6)$ and 13 for $(D_7,D_4)$). 

The second flow to $(A_5,E_6)$ is a bit harder since we need to work out the effect of a (hidden) FI parameter at the $C_2$ node, therefore we make a short digression to discuss this. It will suffice for our purposes to understand this problem for the following family of theories: 
\be\label{step1}
[D_4]-C_2- \U(1)-[k] 
\ee 
where we have $k$ hypermultiplets charged under the $\U(1)$ gauge group. It is easier to approach the problem in the mirror dual theory of \eqref{step1}, where FI parameters are mapped to masses. The mirror dual of \eqref{step1} is given by the following quiver: 
\bes{\label{mirrstep1}
\begin{tikzpicture}[baseline, font=\footnotesize, decoration={brace}]
\node[draw=none] (u3) at (0,0) {$\U(3)$};
\node[draw=none] (u2R) at (2,0) {$\U(2)$};
\node[draw=none] (u1R) at (4,0) {$\U(1)$};
\node[draw=none] (u1R1) at (3.5,0.3) {};
\node[draw=none] (dots) at (6,0) {$\dots$};
\node[draw=none] (u1RR) at (8,0) {$\U(1)$};
\node[draw=none] (u1RR1) at (8.5,0.3) {};
\node[draw=none] (flavR) at (9.5,0) {$[1]$};
\node[draw=none] (u2L) at (-2,-1) {$\U(2)$};
\node[draw=none] (u2LL) at (-2,1) {$\U(2)$};
\node[draw=none] (flavL) at (-3.5,-1) {$[1]$};
\node[draw=none] (flavLL) at (-3.5,1) {$[1]$};
\draw (flavLL)--(u2LL)--(u3)--(u2R)--(u1R)--(dots)--(u1RR)--(flavR);
\draw (flavL)--(u2L)--(u3);
\draw[decorate] (u1R1) -- (u1RR1) node[midway, above=0.2cm] {$k$};
\end{tikzpicture}
}
In the above quiver we have two mass parameters, one can be taken to act on one of the two flavors on the left (and corresponds to the $C_2$ FI parameter) whereas the second acts on the $\U(1)$ flavor on the right and corresponds to the $U(1)$ FI parameter. We can indeed check that symmetries and moduli space dimensions match as expected. Furthermore, if we turn on a mass for the rightmost flavor the abelian nodes become underbalanced and can be dualized to $k$ twisted hypermultiplets, thus matching the effect of ungauging the $\U(1)$ gauge group in \eqref{step1}. 

In order to understand the effect of the (hidden) FI parameter for the balanced $C_2$ in \eqref{step1}, we give mass to one of the flavors on the left in \eqref{mirrstep1}. This makes the corresponding $\U(2)$ node underbalanced, and therefore we can replace it with an abelian node plus a free (twisted) hypermultiplet thanks to the duality \eref{dualityGW}. This then makes the $\U(3)$ node underbalanced, and we can dualize it as well, producing more underbalanced nodes. At the end of the dualization sequence (which involves six steps) all the gauge groups become abelian, and we land on the mirror of SQED with $k+4$ flavors plus six free hypermultiplets. Our claim is therefore that the effect of the FI parameter at the $C_2$ balanced node is to induce the RG flow: 
\be\label{C2FI}
[D_4]-C_2- \U(1)-[k] \;\longrightarrow \;\U(1)-[k+4]\; +\;\text{6 free hypermultiplets}  
\ee 
Armed with this result, it becomes easy to check that turning on the (hidden) FI parameter for the balanced $C_2$ node in \ref{mirrD12m5E6} and also for the abelian node on its left we obtain $\eref{quiv2massesE6}_{\fn=2}$ and 9 free hypermultiplets. Out of these, five account for the difference in rank between the UV and IR theories and the other four fit in the mirror dual of $(A_5, E_6)$.  

\subsubsection*{Test by gauging the baryonic/topological symmetries}
Another non-trivial test of $\eref{mirrD12m5E6}_{\fn=1}$ comes from the observation that, upon reduction to 3d, the $(D_7, E_6)$ yields the following theory
\bes{
\begin{tikzpicture}[baseline, font=\footnotesize]
\node[draw=none] (c) at (0,0) {$\SU(6)$};
\node[draw=none] (su3R) at (2,0) {$\U(4)$};
\node[draw=none] (u1R) at (4,0) {$\SU(3)$};
\node[draw=none] (u1RR) at (6,0) {$\U(1)$};
\node[draw=none] (su3L) at (-2,0) {$\SU(4)$};
\node[draw=none] (u1L) at (-4,0) {$\SU(2)$};
\node[draw=none] (su3B) at (0,1.5) {$\U(3)$};
\node[draw=none] (u1B) at (0,3) {$\SU(1)$};
\draw (u1L)--(su3L)--(c)--(su3R)--(u1R)--(u1RR);
\draw (c)--(su3B)--(u1B);
\end{tikzpicture}
}
Suppose that we turn all special unitary gauge groups $\SU(n)$ in the above quiver into unitary gauge groups $\U(n)$ by gauging the baryonic symmetries. We observe that certain gauge groups are underbalanced and upon using the duality \eref{dualityGW} repeatedly, we obtain
\bes{ \label{starshapedE6w12twisted}
&\begin{tikzpicture}[baseline, font=\footnotesize]
\node[draw=none] (c) at (0,0) {$\U(3)$};
\node[draw=none] (su3R) at (2,0) {$\U(2)$};
\node[draw=none] (u1R) at (4,0) {$\U(1)$};
\node[draw=none] (su3L) at (-2,0) {$\U(2)$};
\node[draw=none] (u1L) at (-4,0) {$\U(1)$};
\node[draw=none] (su3B) at (0,1.5) {$\U(2)$};
\node[draw=none] (u1B) at (0,3) {$\U(1)$};
\draw (u1L)--(su3L)--(c)--(su3R)--(u1R);
\draw (c)--(su3B)--(u1B);
\end{tikzpicture} \\
& \text{+ 12 twisted hypermultiplets}
}
Note that the above star-shaped quiver is the mirror theory for the 4d rank-one Minahan-Nemeschansky (MN) $E_6$ theory. Let us now derive the mirror theory of this from the proposal $\eref{mirrD12m5E6}_{\fn=1}$.  In the view of the mirror theory, the aforementioned procedure of turning $\SU(n)$ into $\U(n)$ corresponds to gauging the topological symmetry $\U(1)^4$ in $\eref{mirrD12m5E6}_{\fn=1}$. We turn all $\U(1)$ gauge nodes into flavor nodes and gauge the emergent $\U(1)$ topological symmetry coming from the balanced $C_2$ node.  As a result, we obtain $12$ free hypermultiplets, together with the $\USp(4)$ gauge theory with $5$ flavors, with the emergent topological symmetry arising from the balanced $\USp(4)$ gauge group being gauged.  The latter is identified as the 3d reduction of the rank-one MN $E_6$ theory \cite[(8.4)]{Carta:2021whq}. We thus obtain the mirror theory of \eref{starshapedE6w12twisted}, as required.

\subsubsection{0 mass parameter}
Any $(D_n, E_6)$ theory such that $n\neq 7 \, (\mod \,\, 12)$ has zero mass parameter. We have discussed the $(D_3, E_6) \cong (A_3, E_6)$ theory in Section \ref{sec:mirrE6mass0}. In the rest of the section, we discuss the $(D_4, E_6)$ theory.

Several properties of the $(D_4, E_6)$ theories have been studied in \cite{Closset:2020afy}, where it was denoted by $E_6^{(1)}$ in that reference. It was shown that this theory is ``bad'' in the sense that its dimensional reduction to 3d does not flow in the IR to a conventional superconformal fixed point. To cure this badness, the authors of \cite{Closset:2020afy} proposed a different definition of the theory, by realizing it as the higgsing of a good theory with a conventional IR fixed point. This is analogous to the procedure proposed in \cite{Gaiotto:2012uq}. It turns out that there are two choices for the higgsing: One is to the (3d mirror of) rank-one $E_6$ theory with a free hypermultiplet, and the other is to a theory of 12 free hypermultiplets. This ambiguity reflects the badness of the theory. Let us explain this point in detail.

It was pointed out in \cite{Closset:2020afy} that the $(D_4, E_6)$ theory dimensionally-reduced to 3d is described by
\bes{
&\begin{tikzpicture}[baseline, font=\footnotesize]
\node[draw=none] (c) at (0,0) {$\SU(4)$};
\node[draw=none] (su3R) at (2,0) {$\U(2)$};
\node[draw=none] (u1R) at (4,0) {$\U(1)$};
\node[draw=none] (su3L) at (-2,0) {$\U(2)$};
\node[draw=none] (u1L) at (-4,0) {$\U(1)$};
\node[draw=none] (su3B) at (0,1.5) {$\U(2)$};
\node[draw=none] (u1B) at (0,3) {$\U(1)$};
\draw (u1L)--(su3L)--(c)--(su3R)--(u1R);
\draw (c)--(su3B)--(u1B);
\end{tikzpicture} 
}
We can consider another quiver that differs from the above by a $\BZ_4$ discrete quotient, reflecting the presence of the $\BZ_4$ 1-form symmetry in the $(D_4, E_6)$ theory:
\bes{\label{badmq}
&\begin{tikzpicture}[baseline, font=\footnotesize]
\node[draw=none] (c) at (0,0) {$\U(4)$};
\node[draw=none] (su3R) at (2,0) {$\U(2)$};
\node[draw=none] (u1R) at (4,0) {$\U(1)$};
\node[draw=none] (su3L) at (-2,0) {$\U(2)$};
\node[draw=none] (u1L) at (-4,0) {$\U(1)$};
\node[draw=none] (su3B) at (0,1.5) {$\U(2)$};
\node[draw=none] (u1B) at (0,3) {$\U(1)$};
\draw (u1L)--(su3L)--(c)--(su3R)--(u1R);
\draw (c)--(su3B)--(u1B);
\end{tikzpicture} 
}
This theory is bad, since it contains monopole operators which violate the unitarity bound. 
We can observe that the ambiguity in the higgsing mentioned before is compatible with the findings of \cite{Assel:2017jgo}: The gauge node at the center of 
the quiver \eref{badmq} is $\U(4)$ with 6 flavors. In \cite{Assel:2017jgo} it was shown that the effective low-energy theory at the most singular point in the Coulomb branch is $\U(3)$ with 6 flavors plus a free (twisted) hypermultiplet. With this substitution, \eref{badmq} becomes the mirror dual of the $E_6$ theory plus a free hyper, in perfect agreement with the above proposal. On the other hand, the Coulomb branch of $U(4)$ with 6 flavors also includes another singular point, whose low-energy effective theory is $\U(2)$ with 6 flavors plus two twisted hypermultiplets (see also \cite{Yaakov:2013fza}). By replacing in \eref{badmq} the $\U(4)$ at the center with a $\U(2)$ and use duality \eref{dualityGW} repeatedly, we find an ugly theory equivalent to a collection of 12 twisted hypermultiplets, in agreement with our previous discussion.

\subsection{Some results on $(D_n, E_7)$ and $(D_n, E_8)$ theories}

For the $(D_n, E_7)$ theories, the number of crepant divisors is $1$ for $n=4,5$;  $2$ for $6\leq n\leq 9$; and then $3$ for $n\geq 10$.  Each of these theories has either $8$, $2$ or $1$ mass parameters.  We find that the $(D_{10}, E_7)$ theory, which has 8 mass parameters, admits the following Lagrangian description:
\bes{
\begin{tikzpicture}[baseline, font=\footnotesize]
\node[draw=none](c) at (0,0) {$\SU(9)$};
\node[draw=none](L1) at (-2,0) {$\SU(6)$};
\node[draw=none](L2) at (-4,0) {$\SU(3)$};
\node[draw=none](R1) at (2,0) {$\SU(7)$};
\node[draw=none](R2) at (4,0) {$\SU(5)$};
\node[draw=none](R3) at (6,0) {$\SU(3)$};
\node[draw=none](R4) at (8,0) {$\SU(1)$};
\node[draw=none](T1) at (0,1.5) {$\SU(5)$};
\node[draw=none](T2) at (0,3) {$\SU(1)$};
\draw (L2)--(L1)--(c)--(R1)--(R2)--(R3)--(R4);
\draw (c)--(T1)--(T2);
\end{tikzpicture}
}

For the $(D_n, E_8)$ theories, the number of crepant divisors is $1$ for $n=4,5$; $2$ for $6\geq n \geq 9$; $3$ for $10 \geq n\geq 15$; and $4$ for $n \geq 16$.  Each of these theories has either $8$ or $0$ mass parameters. We find that the $(D_{16}, E_8)$, which has $8$ mass parameters, admits the following Lagrangian description:
\bes{
\begin{tikzpicture}[baseline, font=\footnotesize]
\node[draw=none](c) at (0,0) {$\SU(15)$};
\node[draw=none](L1) at (-2,0) {$\SU(10)$};
\node[draw=none](L2) at (-4,0) {$\SU(5)$};
\node[draw=none](R1) at (2,0) {$\SU(12)$};
\node[draw=none](R2) at (4,0) {$\SU(9)$};
\node[draw=none](R3) at (6,0) {$\SU(6)$};
\node[draw=none](R4) at (8,0) {$\SU(3)$};
\node[draw=none](T1) at (0,1.5) {$\SU(8)$};
\node[draw=none](T2) at (0,3) {$\SU(1)$};
\draw (L2)--(L1)--(c)--(R1)--(R2)--(R3)--(R4);
\draw (c)--(T1)--(T2);
\end{tikzpicture}
}

\acknowledgments

We thank Mario De Marco, Craig Lawrie, Nakarin Lohitsiri, Andrea Sangiovanni and Roberto Valandro for valuable discussions. F. C. is supported by STFC consolidated grant ST/T000708/1. The work of S. G. is supported by the INFN grant ``Per attività di formazione per sostenere progetti di ricerca'' (GRANT 73/STRONGQFT). The work of A. M. is supported in part by Deutsche Forschungsgemeinschaft under Germany's Excellence Strategy EXC 2121 Quantum Universe 390833306.

\appendix

\section{Notation and convention}
\label{app:conv}

Consistently with the notation adopted in \cite{Carta:2021whq,Carta:2021dyx}, we will call $\SO(2N) = D_N$, $\USp(2N) = C_N$ and $\SO(2N + 1) = B_N$. We denote the diagonal $\BZ_2$ and $\U(1)$ quotient of the gauge symmetry by $/\BZ_2$ and $/\U(1)$, respectively.

As introduced in \cite{Gaiotto:2008ak}, a $\USp(2N)$ gauge group with $N_f$ fundamental flavors in a $3$d $\CN=4$ gauge theory is balanced if $N_f=2N+1$. If the number of flavors are less (more) than $2N+1$ the group is underbalanced (overbalanced). Moreover, the condition for $\USp(2N)$ to have a zero beta-function is $N_f=2N+2$, i.e. in $3$d theory, the $\USp(2N)$ group is overbalanced. A similar discussion can be done also for $\SO(N)$ gauge groups with $N_f$ fundamental flavors for which the balancing condition is obtained for $N_f=N-1$, while a zero beta-function requires $N_f=N-2$, i.e. underbalanced in $3$d.

We adopt the following notation for the quiver diagrams:
\bi
\item We denote $M$ copies of half-hypermultiplets in the representation $[\mathbf{2}; \mathbf{2}]$ of the gauge group $\SO(2) \times \SO(2)$ by
\bes{ \label{blueedge}
D_1 \begin{tikzpicture}[baseline] \draw[draw,solid,blue,thick] (0,0.1)--(1,0.1) node[midway, above] {\blue \scriptsize $M$}; \end{tikzpicture} D_1\fstop
}
It gives rise to a $\U(M)^2/\U(1)$ flavor symmetry, whose algebra is isomorphic to $\SU(M) \times \SU(M) \times \U(1)$. To make the Cartan elements of the latter manifest, we should interpret \eref{blueedge} as denoting the half-hypermultiplets in the following representation of $\left\{\U(1) \times \U(1)\right\} \times \SU(M) \times \SU(M) \times \U(1)$, where each of the first two $\U(1)$ factors are isomorphic to each $\SO(2)$ gauge group:
\bes{ \label{repblueedge}
&[+1; +1; \bar{\mathbf{M}}; \mathbf{1}; -1] \oplus [-1; -1; \mathbf{M}; \mathbf{1}; +1 ] \\
&\oplus [+1; -1;  \mathbf{1}; \mathbf{M}; +1]\oplus [-1; +1;  \mathbf{1}; \bar{\mathbf{M}}; -1]~.  
}

\item The hypermultiplets carrying charge $2$ under $\U(1)\cong \SO(2)$ is denoted by a zigzag line and by a subscript $2$:
\bes{ \label{wiggleline}
D_1  \begin{tikzpicture}[baseline] \draw[draw,solid,black,snake it] (0,0.1)--(1,0.1) node[midway, above] {}; \end{tikzpicture} [F]_2\fstop
}
 This gives rise to a $\SU(F)$ flavor symmetry. 
\item A black line between $G_1$ and $G_2$ denotes matter in the bifundamental representation of $G_1 \times G_2$, as usual.
\ei

\section{Coulomb branch Hilbert series of \eqref{redA5A6}}
\label{app:3dA5E6CBHS}

Let us consider \eqref{redA5A6} and we associate to each node a magnetic flux variable that we can use to compute the CB Hilbert series, as follows
\bes{ \label{eq:A5E6fug}
\begin{tikzpicture}[baseline,font=\footnotesize]
\node[draw=none] (U1) at (-4,0) {$\U(1)$};
\node[draw=none] (U2) at (-2,0) {$\U(2)$};
\node[draw=none] (SU4) at (0,0) {$\SU(4)$};
\node[draw=none] (U3) at (2,0) {$\U(3)$};
\node[draw=none] (U2p) at (4,0) {$\U(2)$};
\node[draw=none] (SU2) at (6,0) {$\SU(2)$};
\node[draw=none] (U1p) at (8,0) {$\U(1)$};
\node[draw=none] (SU2p) at (0,1.5) {$\SU(2)$};
\node[draw=none]  at (-4,-0.6) {$m^{\U(1)}$};
\node[draw=none]  at (-2,-0.6) {$m^{\U(2)}_{1,2}$};
\node[draw=none]  at (0,-0.6) {$m^{\SU(4)}_{1,2,3,4}$};
\node[draw=none]  at (2,-0.6) {$m^{\U(3)}_{1,2,3}$};
\node[draw=none]  at (4,-0.6) {$n^{\U(2)}_{1,2}$};
\node[draw=none]  at (6,-0.6) {$n^{\SU(2)}_{1,2}$};
\node[draw=none]  at (8,-0.6) {$n^{\U(1)}$};
\node[draw=none] at (0,2.1) {$m^{\SU(2)}_{1,2}$};
\draw (U1)--(U2)--(SU4)--(U3)--(U2p)--(SU2)--(U1p);
\draw (SU4)--(SU2p);
\end{tikzpicture}
}
Recall that the $\SU(N)$ magnetic fugacities are constrained to sum to zero. To perform the computation, it is simpler to break up the computation in to three parts corresponding to the left, right, and top arms of the quiver, and then ``glue'' the three contributions together at the central $\SU(4)$ node.  

The contribution of the left arm reads
\bes{
H_L(t; m^{\SU(4)}_{1,2,3,4}) = \sum_{m^{\U(1)}\in \BZ} \,\, \sum_{m^{\U(2)}_1\geq m^{\U(2)}_2> -\infty} t^{\Delta_L} P_{\U(1)}(m^{\U(1)})P_{\U(2)}(\vec m^{\U(2)})~,
}
where the function $P_G(\vec m)$ is given by \cite[Appendix A]{Cremonesi:2013lqa} and
\bes{
\Delta_L = \frac{1}{2} \left( \sum_{i=1}^2\left| m^{\U(1)}- m^{\U(2)}_i\right| +  \sum_{i=1}^2\sum_{j=1}^4\left| m^{\U(2)}_i- m^{\SU(4)}_j\right| \right)-\left| m^{\U(2)}_1-m^{\U(2)}_2 \right| ~.
}
The contribution of the top arm reads
\bes{
&H_T(t; m^{\SU(4)}_{1,2,3,4}) \\
&=  \sum_{m^{\SU(2)}_1\geq m^{\SU(2)}_2> -\infty} t^{\Delta_T} P_{\U(2)}\left(\vec{m}^{\SU(2)} \right) (1-t) \times \delta(m^{\SU(2)}_1+ m^{\SU(2)}_2)~,
}
where
\bes{
\Delta_T = \frac{1}{2} \sum_{i=1}^2\sum_{j=1}^4 \left| m^{\SU(2)}_i- m^{\SU(4)}_j\right|  -\left| m^{\SU(2)}_1-m^{\SU(2)}_2 \right|~.
}
The contribution of the right arm reads
\bes{
& H_R(t; m^{\SU(4)}_{1,2,3,4}) \\
&=\sum_{m^{\U(3)}_2\geq m^{\U(3)}_3> -\infty} \,\, \sum_{n^{\U(2)}_1\geq n^{\U(2)}_2> -\infty}\,\, \sum_{n^{\SU(2)}_1\geq n^{\SU(2)}_2> -\infty} \,\, \sum_{n^{\U(1)} \in \BZ} t^{\Delta_R} \times P_{\U(3)}\left(\vec{m}^{\U(3)} \right)\times  \\
&\qquad  P_{\U(2)}\left(\vec{n}^{\U(2)} \right) P_{\U(2)}\left(\vec{n}^{\SU(2)} \right) (1-t) P_{\U(1)}\left(n^{\U(1)} \right) \times \delta(n^{\SU(2)}_1+ n^{\SU(2)}_2)~,
}
where
\bes{
\scalebox{0.8}{$
\begin{split}
\Delta_R  = \frac{1}{2}& \left(  + \sum_{i=1}^3\sum_{j=1}^4\left| m^{\U(3)}_i- m^{\SU(4)}_j\right|  + \sum_{i=1}^3\sum_{j=1}^2\left| m^{\U(3)}_i- n^{\U(2)}_j\right| + \sum_{i=1}^2\sum_{j=1}^2\left| n^{\U(2)}_i- n^{\SU(2)}_j\right| +\right. \\
        &\left. \, + \sum_{i=1}^2\left| n^{\SU(2)}_i- n^{\U(1)}\right|\right) - \left(  \sum_{i=1}^3\sum_{j=i+1}^3\left| m^{\U(3)}_i- m^{\U(3)}_j\right| + \sum_{i=1}^2\sum_{j=i+1}^2\left| n^{\U(2)}_i- n^{\U(2)}_j\right| +\right.  \\
        &\left. \,+ \sum_{i=1}^2\sum_{j=i+1}^2\left| n^{\SU(2)}_i- n^{\SU(2)}_j\right|\right)~.
\end{split}$  }      
}

The (unrefined) Coulomb branch Hilbert series of \eref{eq:A5E6fug} with an overall $\BZ_2$ quotient is then given by
\bes{
&H_{\text{\eref{eq:A5E6fug}}/\BZ_2} (t ) \\
&= \sum_{\epsilon=0}^1 \,\, \sum_{\substack{m^{\SU(4)}_1\geq \cdots \geq m^{\SU(4)}_4> -\infty\\m^{\SU(4)}_i \in \BZ + \frac{1}{2} \epsilon}}  \delta\left( \sum_{i=4} m^{\SU(4)}_i \right)t^{-\sum_{i=1}^4\sum_{j=i+1}^4\left| m^{\SU(4)}_i- m^{\SU(4)}_j\right|} \times \\
& \qquad H_L(t; m^{\SU(4)}_{1,2,3,4}) H_T(t; m^{\SU(4)}_{1,2,3,4}) H_R(t; m^{\SU(4)}_{1,2,3,4}) \times P_{\U(4)}(\vec m^{\SU(4)}) (1-t) ~.
}
We remark that the integral-valued magnetic fluxes $(m^{\SU(4)}_1, \cdots, {m^{\SU(4)}_4})$ that contribute to the above expression up to order $t^{3/2}$ are
\bes{
(0, 0, 0, 0), \, (1, 0, 0, -1), \, (1, 1, -1, -1), \, (1, 1, 0, -2),\, (2, 0, -1, -1)~,
}
which gives 
\bes{
1 + 58 t + 256 t^{3/2} + \ldots~,
}
and the half-odd-integral-valued magnetic fluxes $({m^{\SU(4)}_1, \cdots, m^{\SU(4)}_4})$ that contribute to the above expression up to order $t^{3/2}$ are
\bes{
\scalebox{0.8}{$
\begin{split}
\left(\frac{1}{2},\frac{1}{2},-\frac{1}{2},-\frac{1}{2}\right),\, \left(\frac{1}{2},\frac{1}{2},\frac{1}{2},-\frac{3}{2}\right),\, \left(\frac{3}{2},-\frac{1}{2},-\frac{1}{2},-\frac{1}{2}\right),\, \left(\frac{3}{2},\frac{1}{2},-\frac{1}{2},-\frac{3}{2}\right),\, \left(\frac{3}{2},\frac{3}{2},-\frac{3}{2},-\frac{3}{2}\right)
\end{split}$}
}
which gives
\bes{
8 t^{1/2} + 24 t + 360 t^{3/2} + \ldots~.
}
Summing these two contributions up, we obtain
\bes{
&1 + 8 t^{1/2} + 82 t + 616 t^{3/2} + \ldots  \\
&= (1-t^{1/2})^{-8} (1 + 46 t + 128 t^{3/2}+ \ldots)~.
}

\section{Non-Higgsable SCFTs with 1-form symmetries}
\label{app:nHSCFTs1form}

In this Appendix, we list many non-Higgsable SCFTs of the type $(G, G')$ with $0 \leq 24(c - a) < 1$, along with their ranks and 1-form symmetries. The data have been computed using a modified code provided in \cite{Carta:2020plx}.\footnote{It is possible to reproduce the same results also using the code provided in \cite{Closset:2021lwy}.}   We omit the $(A_n, A_m)$ theories from this list since all of them have a trivial 1-form symmetry and their information has already been provided in \cite[Appendix C]{Carta:2021whq}.

\begin{center}
\renewcommand{\arraystretch}{1.25}
\begin{longtable}{c|c|c|c||c|c|c|c||c|c|c|c}
\caption{$(A_n,D_m)$ with $1\leq n,m\leq 100$. We called $h=24(c-a)$, $r$ the rank of the SCFT and $s$ the $1$-form symmetry group. We list only those theories with $0\leq h < 1$.}\\
$(A_n,D_m)$  & $h$ & $r$ & $s$ & $(A_n,D_m)$  & $h$ & $r$ & $s$ & $(A_n,D_m)$  & $h$ & $r$ & $s$\\
    \hline
\endhead
\label{tab:AnDmNHC}$(A_2,D_3)$     & $\frac{3}{7}$    & $3  $ & $1$            & $(A_2,D_{87})$  & $\frac{87}{175}$  & $87 $  & $1$              & $(A_4,D_{73})$     & $\frac{146}{149}$ & $146 $ & $1$ \\
$(A_2,D_4)$     & $0$              & $4  $ & $\mathbb{Z}_2$ & $(A_2,D_{88})$  & $\frac{28}{59}$   & $88 $  & $\mathbb{Z}_2$   & $(A_4,D_{74})$     & $\frac{148}{151}$ & $148 $ & $1$                 \\
$(A_2,D_5)$     & $\frac{5}{11}$   & $5  $ & $1$            & $(A_2,D_{89})$  & $\frac{89}{179}$  & $89 $  & $1$              & $(A_4,D_{75})$     & $\frac{50}{51}$   & $150 $ & $1$                 \\
$(A_2,D_6)$     & $\frac{6}{13}$   & $6  $ & $1$            & $(A_2,D_{90})$  & $\frac{90}{181}$  & $90 $  & $1$              & $(A_4,D_{76})$     & $\frac{28}{31}$   & $152 $ & $\mathbb{Z}_2^2$    \\
$(A_2,D_7)$     & $\frac{1}{5}$    & $7  $ & $\mathbb{Z}_2$ & $(A_2,D_{91})$  & $\frac{29}{61}$   & $91 $  & $\mathbb{Z}_2$   & $(A_4,D_{77})$     & $\frac{154}{157}$ & $154 $ & $1$                 \\
$(A_2,D_8)$     & $\frac{8}{17}$   & $8  $ & $1$            & $(A_2,D_{92})$  & $\frac{92}{185}$  & $92 $  & $1$              & $(A_4,D_{78})$     & $\frac{52}{53}$   & $156 $ & $1$                 \\
$(A_2,D_9)$     & $\frac{9}{19}$   & $9  $ & $1$            & $(A_2,D_{93})$  & $\frac{93}{187}$  & $93 $  & $1$              & $(A_4,D_{79})$     & $\frac{158}{161}$ & $158 $ & $1$                 \\
$(A_2,D_{10})$  & $\frac{2}{7}$    & $10 $ & $\mathbb{Z}_2$ & $(A_2,D_{94})$  & $\frac{10}{21}$   & $94 $  & $\mathbb{Z}_2$   & $(A_4,D_{80})$     & $\frac{160}{163}$ & $160 $ & $1$                 \\
$(A_2,D_{11})$  & $\frac{11}{23}$  & $11 $ & $1$            & $(A_2,D_{95})$  & $\frac{95}{191}$  & $95 $  & $1$              & $(A_4,D_{81})$     & $\frac{10}{11}$   & $162 $ & $\mathbb{Z}_2^2$    \\
$(A_2,D_{12})$  & $\frac{12}{25}$  & $12 $ & $1$            & $(A_2,D_{96})$  & $\frac{96}{193}$  & $96 $  & $1$              & $(A_4,D_{82})$     & $\frac{164}{167}$ & $164 $ & $1$                 \\
$(A_2,D_{13})$  & $\frac{1}{3}$    & $13 $ & $\mathbb{Z}_2$ & $(A_2,D_{97})$  & $\frac{31}{65}$   & $97 $  & $\mathbb{Z}_2$   & $(A_4,D_{83})$     & $\frac{166}{169}$ & $166 $ & $1$                 \\
$(A_2,D_{14})$  & $\frac{14}{29}$  & $14 $ & $1$            & $(A_2,D_{98})$  & $\frac{98}{197}$  & $98 $  & $1$              & $(A_4,D_{84})$     & $\frac{56}{57}$   & $168 $ & $1$                 \\
$(A_2,D_{15})$  & $\frac{15}{31}$  & $15 $ & $1$            & $(A_2,D_{99})$  & $\frac{99}{199}$  & $99 $  & $1$              & $(A_4,D_{85})$     & $\frac{170}{173}$ & $170 $ & $1$                 \\
$(A_2,D_{16})$  & $\frac{4}{11}$   & $16 $ & $\mathbb{Z}_2$ & $(A_2,D_{100})$ & $\frac{32}{67}$   & $100$  & $\mathbb{Z}_2$   & $(A_4,D_{86})$     & $\frac{32}{35}$   & $172 $ & $\mathbb{Z}_2^2$    \\
$(A_2,D_{17})$  & $\frac{17}{35}$  & $17 $ & $1$            & $(A_4,D_3)$     & $\frac{2}{3}$     & $6  $  & $1$              & $(A_4,D_{87})$     & $\frac{58}{59}$   & $174 $ & $1$                 \\
$(A_2,D_{18})$  & $\frac{18}{37}$  & $18 $ & $1$            & $(A_4,D_4)$     & $\frac{8}{11}$    & $8  $  & $1$              & $(A_4,D_{88})$     & $\frac{176}{179}$ & $176 $ & $1$                 \\
$(A_2,D_{19})$  & $\frac{5}{13}$   & $19 $ & $\mathbb{Z}_2$ & $(A_4,D_5)$     & $\frac{10}{13}$   & $10 $  & $1$              & $(A_4,D_{89})$     & $\frac{178}{181}$ & $178 $ & $1$                 \\
$(A_2,D_{20})$  & $\frac{20}{41}$  & $20 $ & $1$            & $(A_4,D_6)$     & $0$               & $12 $  & $\mathbb{Z}_2^2$ & $(A_4,D_{90})$     & $\frac{60}{61}$   & $180 $ & $1$                 \\
$(A_2,D_{21})$  & $\frac{21}{43}$  & $21 $ & $1$            & $(A_4,D_7)$     & $\frac{14}{17}$   & $14 $  & $1$              & $(A_4,D_{91})$     & $\frac{34}{37}$   & $182 $ & $\mathbb{Z}_2^2$    \\
$(A_2,D_{22})$  & $\frac{2}{5}$    & $22 $ & $\mathbb{Z}_2$ & $(A_4,D_8)$     & $\frac{16}{19}$   & $16 $  & $1$              & $(A_4,D_{92})$     & $\frac{184}{187}$ & $184 $ & $1$                 \\
$(A_2,D_{23})$  & $\frac{23}{47}$  & $23 $ & $1$            & $(A_4,D_9)$     & $\frac{6}{7}$     & $18 $  & $1$              & $(A_4,D_{93})$     & $\frac{62}{63}$   & $186 $ & $1$                 \\
$(A_2,D_{24})$  & $\frac{24}{49}$  & $24 $ & $1$            & $(A_4,D_{10})$  & $\frac{20}{23}$   & $20 $  & $1$              & $(A_4,D_{94})$     & $\frac{188}{191}$ & $188 $ & $1$                 \\
$(A_2,D_{25})$  & $\frac{7}{17}$   & $25 $ & $\mathbb{Z}_2$ & $(A_4,D_{11})$  & $\frac{2}{5}$     & $22 $  & $\mathbb{Z}_2^2$ & $(A_4,D_{95})$     & $\frac{190}{193}$ & $190 $ & $1$                 \\
$(A_2,D_{26})$  & $\frac{26}{53}$  & $26 $ & $1$            & $(A_4,D_{12})$  & $\frac{8}{9}$     & $24 $  & $1$              & $(A_4,D_{96})$     & $\frac{12}{13}$   & $192 $ & $\mathbb{Z}_2^2$    \\
$(A_2,D_{27})$  & $\frac{27}{55}$  & $27 $ & $1$            & $(A_4,D_{13})$  & $\frac{26}{29}$   & $26 $  & $1$              & $(A_4,D_{97})$     & $\frac{194}{197}$ & $194 $ & $1$                 \\
$(A_2,D_{28})$  & $\frac{8}{19}$   & $28 $ & $\mathbb{Z}_2$ & $(A_4,D_{14})$  & $\frac{28}{31}$   & $28 $  & $1$              & $(A_4,D_{98})$     & $\frac{196}{199}$ & $196 $ & $1$                 \\
$(A_2,D_{29})$  & $\frac{29}{59}$  & $29 $ & $1$            & $(A_4,D_{15})$  & $\frac{10}{11}$   & $30 $  & $1$              & $(A_4,D_{99})$     & $\frac{66}{67}$   & $198 $ & $1$                 \\
$(A_2,D_{30})$  & $\frac{30}{61}$  & $30 $ & $1$            & $(A_4,D_{16})$  & $\frac{4}{7}$     & $32 $  & $\mathbb{Z}_2^2$ & $(A_4,D_{100})$    & $\frac{200}{203}$ & $200 $ & $1$                 \\
$(A_2,D_{31})$  & $\frac{3}{7}$    & $31 $ & $\mathbb{Z}_2$ & $(A_4,D_{17})$  & $\frac{34}{37}$   & $34 $  & $1$              & $(A_6,D_3)$        & $\frac{9}{11}$    & $9   $ & $1$                 \\
$(A_2,D_{32})$  & $\frac{32}{65}$  & $32 $ & $1$            & $(A_4,D_{18})$  & $\frac{12}{13}$   & $36 $  & $1$              & $(A_6,D_4)$        & $\frac{12}{13}$   & $12  $ & $1$                 \\
$(A_2,D_{33})$  & $\frac{33}{67}$  & $33 $ & $1$            & $(A_4,D_{19})$  & $\frac{38}{41}$   & $38 $  & $1$              & $(A_6,D_8)$        & $0$               & $24  $ & $\mathbb{Z}_2^3$    \\
$(A_2,D_{34})$  & $\frac{10}{23}$  & $34 $ & $\mathbb{Z}_2$ & $(A_4,D_{20})$  & $\frac{40}{43}$   & $40 $  & $1$              & $(A_6,D_{15})$     & $\frac{3}{5}$     & $45  $ & $\mathbb{Z}_2^3$    \\
$(A_2,D_{35})$  & $\frac{35}{71}$  & $35 $ & $1$            & $(A_4,D_{21})$  & $\frac{2}{3}$     & $42 $  & $\mathbb{Z}_2^2$ & $(A_6,D_{22})$     & $\frac{6}{7}$     & $66  $ & $\mathbb{Z}_2^3$    \\
$(A_2,D_{36})$  & $\frac{36}{73}$  & $36 $ & $1$            & $(A_4,D_{22})$  & $\frac{44}{47}$   & $44 $  & $1$              & $(A_8,D_3)$        & $\frac{12}{13}$   & $12  $ & $1$                 \\
$(A_2,D_{37})$  & $\frac{11}{25}$  & $37 $ & $\mathbb{Z}_2$ & $(A_4,D_{23})$  & $\frac{46}{49}$   & $46 $  & $1$              & $(A_8,D_4)$        & $\frac{4}{5}$     & $16  $ & $\mathbb{Z}_2$      \\
$(A_2,D_{38})$  & $\frac{38}{77}$  & $38 $ & $1$            & $(A_4,D_{24})$  & $\frac{16}{17}$   & $48 $  & $1$              & $(A_8,D_{10})$     & $0$               & $40  $ & $\mathbb{Z}_2^4$    \\
$(A_2,D_{39})$  & $\frac{39}{79}$  & $39 $ & $1$            & $(A_4,D_{25})$  & $\frac{50}{53}$   & $50 $  & $1$              & $(A_8,D_{19})$     & $\frac{4}{5}$     & $76  $ & $\mathbb{Z}_2^4$    \\
$(A_2,D_{40})$  & $\frac{4}{9}$    & $40 $ & $\mathbb{Z}_2$ & $(A_4,D_{26})$  & $\frac{8}{11}$    & $52 $  & $\mathbb{Z}_2^2$ & $(A_{10},D_{12})$  & $0$               & $60  $ & $\mathbb{Z}_2^5$    \\
$(A_2,D_{41})$  & $\frac{41}{83}$  & $41 $ & $1$            & $(A_4,D_{27})$  & $\frac{18}{19}$   & $54 $  & $1$              & $(A_{12},D_{14})$  & $0$               & $84  $ & $\mathbb{Z}_2^6$    \\
$(A_2,D_{42})$  & $\frac{42}{85}$  & $42 $ & $1$            & $(A_4,D_{28})$  & $\frac{56}{59}$   & $56 $  & $1$              & $(A_{14},D_{16})$  & $0$               & $112 $ & $\mathbb{Z}_2^7$    \\
$(A_2,D_{43})$  & $\frac{13}{29}$  & $43 $ & $\mathbb{Z}_2$ & $(A_4,D_{29})$  & $\frac{58}{61}$   & $58 $  & $1$              & $(A_{16},D_{18})$  & $0$               & $144 $ & $\mathbb{Z}_2^8$    \\
$(A_2,D_{44})$  & $\frac{44}{89}$  & $44 $ & $1$            & $(A_4,D_{30})$  & $\frac{20}{21}$   & $60 $  & $1$              & $(A_{18},D_{20})$  & $0$               & $180 $ & $\mathbb{Z}_2^9$    \\
$(A_2,D_{45})$  & $\frac{45}{91}$  & $45 $ & $1$            & $(A_4,D_{31})$  & $\frac{10}{13}$   & $62 $  & $\mathbb{Z}_2^2$ & $(A_{20},D_{22})$  & $0$               & $220 $ & $\mathbb{Z}_2^{10}$ \\
$(A_2,D_{46})$  & $\frac{14}{31}$  & $46 $ & $\mathbb{Z}_2$ & $(A_4,D_{32})$  & $\frac{64}{67}$   & $64 $  & $1$              & $(A_{22},D_{24})$  & $0$               & $264 $ & $\mathbb{Z}_2^{11}$ \\
$(A_2,D_{47})$  & $\frac{47}{95}$  & $47 $ & $1$            & $(A_4,D_{33})$  & $\frac{22}{23}$   & $66 $  & $1$              & $(A_{24},D_{26})$  & $0$               & $312 $ & $\mathbb{Z}_2^{12}$ \\
$(A_2,D_{48})$  & $\frac{48}{97}$  & $48 $ & $1$            & $(A_4,D_{34})$  & $\frac{68}{71}$   & $68 $  & $1$              & $(A_{26},D_{28})$  & $0$               & $364 $ & $\mathbb{Z}_2^{13}$ \\
$(A_2,D_{49})$  & $\frac{5}{11}$   & $49 $ & $\mathbb{Z}_2$ & $(A_4,D_{35})$  & $\frac{70}{73}$   & $70 $  & $1$              & $(A_{28},D_{30})$  & $0$               & $420 $ & $\mathbb{Z}_2^{14}$ \\
$(A_2,D_{50})$  & $\frac{50}{101}$ & $50 $ & $1$            & $(A_4,D_{36})$  & $\frac{4}{5}$     & $72 $  & $\mathbb{Z}_2^2$ & $(A_{30},D_{32})$  & $0$               & $480 $ & $\mathbb{Z}_2^{15}$ \\
$(A_2,D_{51})$  & $\frac{51}{103}$ & $51 $ & $1$            & $(A_4,D_{37})$  & $\frac{74}{77}$   & $74 $  & $1$              & $(A_{32},D_{34})$  & $0$               & $544 $ & $\mathbb{Z}_2^{16}$ \\
$(A_2,D_{52})$  & $\frac{16}{35}$  & $52 $ & $\mathbb{Z}_2$ & $(A_4,D_{38})$  & $\frac{76}{79}$   & $76 $  & $1$              & $(A_{34},D_{36})$  & $0$               & $612 $ & $\mathbb{Z}_2^{17}$ \\
$(A_2,D_{53})$  & $\frac{53}{107}$ & $53 $ & $1$            & $(A_4,D_{39})$  & $\frac{26}{27}$   & $78 $  & $1$              & $(A_{36},D_{38})$  & $0$               & $684 $ & $\mathbb{Z}_2^{18}$ \\
$(A_2,D_{54})$  & $\frac{54}{109}$ & $54 $ & $1$            & $(A_4,D_{40})$  & $\frac{80}{83}$   & $80 $  & $1$              & $(A_{38},D_{40})$  & $0$               & $760 $ & $\mathbb{Z}_2^{19}$ \\
$(A_2,D_{55})$  & $\frac{17}{37}$  & $55 $ & $\mathbb{Z}_2$ & $(A_4,D_{41})$  & $\frac{14}{17}$   & $82 $  & $\mathbb{Z}_2^2$ & $(A_{40},D_{42})$  & $0$               & $840 $ & $\mathbb{Z}_2^{20}$ \\
$(A_2,D_{56})$  & $\frac{56}{113}$ & $56 $ & $1$            & $(A_4,D_{42})$  & $\frac{28}{29}$   & $84 $  & $1$              & $(A_{42},D_{44})$  & $0$               & $924 $ & $\mathbb{Z}_2^{21}$ \\
$(A_2,D_{57})$  & $\frac{57}{115}$ & $57 $ & $1$            & $(A_4,D_{43})$  & $\frac{86}{89}$   & $86 $  & $1$              & $(A_{44},D_{46})$  & $0$               & $1012$ & $\mathbb{Z}_2^{22}$ \\
$(A_2,D_{58})$  & $\frac{6}{13}$   & $58 $ & $\mathbb{Z}_2$ & $(A_4,D_{44})$  & $\frac{88}{91}$   & $88 $  & $1$              & $(A_{46},D_{48})$  & $0$               & $1104$ & $\mathbb{Z}_2^{23}$ \\
$(A_2,D_{59})$  & $\frac{59}{119}$ & $59 $ & $1$            & $(A_4,D_{45})$  & $\frac{30}{31}$   & $90 $  & $1$              & $(A_{48},D_{50})$  & $0$               & $1200$ & $\mathbb{Z}_2^{24}$ \\
$(A_2,D_{60})$  & $\frac{60}{121}$ & $60 $ & $1$            & $(A_4,D_{46})$  & $\frac{16}{19}$   & $92 $  & $\mathbb{Z}_2^2$ & $(A_{50},D_{52})$  & $0$               & $1300$ & $\mathbb{Z}_2^{25}$ \\
$(A_2,D_{61})$  & $\frac{19}{41}$  & $61 $ & $\mathbb{Z}_2$ & $(A_4,D_{47})$  & $\frac{94}{97}$   & $94 $  & $1$              & $(A_{52},D_{54})$  & $0$               & $1404$ & $\mathbb{Z}_2^{26}$ \\
$(A_2,D_{62})$  & $\frac{62}{125}$ & $62 $ & $1$            & $(A_4,D_{48})$  & $\frac{32}{33}$   & $96 $  & $1$              & $(A_{54},D_{56})$  & $0$               & $1512$ & $\mathbb{Z}_2^{27}$ \\
$(A_2,D_{63})$  & $\frac{63}{127}$ & $63 $ & $1$            & $(A_4,D_{49})$  & $\frac{98}{101}$  & $98  $ & $1$              & $(A_{56},D_{58})$  & $0$               & $1624$ & $\mathbb{Z}_2^{28}$ \\
$(A_2,D_{64})$  & $\frac{20}{43}$  & $64 $ & $\mathbb{Z}_2$ & $(A_4,D_{50})$  & $\frac{100}{103}$ & $100 $ & $1$              & $(A_{58},D_{60})$  & $0$               & $1740$ & $\mathbb{Z}_2^{29}$ \\
$(A_2,D_{65})$  & $\frac{65}{131}$ & $65 $ & $1$            & $(A_4,D_{51})$  & $\frac{6}{7}$     & $102 $ & $\mathbb{Z}_2^2$ & $(A_{60},D_{62})$  & $0$               & $1860$ & $\mathbb{Z}_2^{30}$ \\
$(A_2,D_{66})$  & $\frac{66}{133}$ & $66 $ & $1$            & $(A_4,D_{52})$  & $\frac{104}{107}$ & $104 $ & $1$              & $(A_{62},D_{64})$  & $0$               & $1984$ & $\mathbb{Z}_2^{31}$ \\
$(A_2,D_{67})$  & $\frac{7}{15}$   & $67 $ & $\mathbb{Z}_2$ & $(A_4,D_{53})$  & $\frac{106}{109}$ & $106 $ & $1$              & $(A_{64},D_{66})$  & $0$               & $2112$ & $\mathbb{Z}_2^{32}$ \\
$(A_2,D_{68})$  & $\frac{68}{137}$ & $68 $ & $1$            & $(A_4,D_{54})$  & $\frac{36}{37}$   & $108 $ & $1$              & $(A_{66},D_{68})$  & $0$               & $2244$ & $\mathbb{Z}_2^{33}$ \\
$(A_2,D_{69})$  & $\frac{69}{139}$ & $69 $ & $1$            & $(A_4,D_{55})$  & $\frac{110}{113}$ & $110 $ & $1$              & $(A_{68},D_{70})$  & $0$               & $2380$ & $\mathbb{Z}_2^{34}$ \\
$(A_2,D_{70})$  & $\frac{22}{47}$  & $70 $ & $\mathbb{Z}_2$ & $(A_4,D_{56})$  & $\frac{20}{23}$   & $112 $ & $\mathbb{Z}_2^2$ & $(A_{70},D_{72})$  & $0$               & $2520$ & $\mathbb{Z}_2^{35}$ \\
$(A_2,D_{71})$  & $\frac{71}{143}$ & $71 $ & $1$            & $(A_4,D_{57})$  & $\frac{38}{39}$   & $114 $ & $1$              & $(A_{72},D_{74})$  & $0$               & $2664$ & $\mathbb{Z}_2^{36}$ \\
$(A_2,D_{72})$  & $\frac{72}{145}$ & $72 $ & $1$            & $(A_4,D_{58})$  & $\frac{116}{119}$ & $116 $ & $1$              & $(A_{74},D_{76})$  & $0$               & $2812$ & $\mathbb{Z}_2^{37}$ \\
$(A_2,D_{73})$  & $\frac{23}{49}$  & $73 $ & $\mathbb{Z}_2$ & $(A_4,D_{59})$  & $\frac{118}{121}$ & $118 $ & $1$              & $(A_{76},D_{78})$  & $0$               & $2964$ & $\mathbb{Z}_2^{38}$ \\
$(A_2,D_{74})$  & $\frac{74}{149}$ & $74 $ & $1$            & $(A_4,D_{60})$  & $\frac{40}{41}$   & $120 $ & $1$              & $(A_{78},D_{80})$  & $0$               & $3120$ & $\mathbb{Z}_2^{39}$ \\
$(A_2,D_{75})$  & $\frac{75}{151}$ & $75 $ & $1$            & $(A_4,D_{61})$  & $\frac{22}{25}$   & $122 $ & $\mathbb{Z}_2^2$ & $(A_{80},D_{82})$  & $0$               & $3280$ & $\mathbb{Z}_2^{40}$ \\
$(A_2,D_{76})$  & $\frac{8}{17}$   & $76 $ & $\mathbb{Z}_2$ & $(A_4,D_{62})$  & $\frac{124}{127}$ & $124 $ & $1$              & $(A_{82},D_{84})$  & $0$               & $3444$ & $\mathbb{Z}_2^{41}$ \\
$(A_2,D_{77})$  & $\frac{77}{155}$ & $77 $ & $1$            & $(A_4,D_{63})$  & $\frac{42}{43}$   & $126 $ & $1$              & $(A_{84},D_{86})$  & $0$               & $3612$ & $\mathbb{Z}_2^{42}$ \\
$(A_2,D_{78})$  & $\frac{78}{157}$ & $78 $ & $1$            & $(A_4,D_{64})$  & $\frac{128}{131}$ & $128 $ & $1$              & $(A_{86},D_{88})$  & $0$               & $3784$ & $\mathbb{Z}_2^{43}$ \\
$(A_2,D_{79})$  & $\frac{25}{53}$  & $79 $ & $\mathbb{Z}_2$ & $(A_4,D_{65})$  & $\frac{130}{133}$ & $130 $ & $1$              & $(A_{88},D_{90})$  & $0$               & $3960$ & $\mathbb{Z}_2^{44}$ \\
$(A_2,D_{80})$  & $\frac{80}{161}$ & $80 $ & $1$            & $(A_4,D_{66})$  & $\frac{8}{9}$     & $132 $ & $\mathbb{Z}_2^2$ & $(A_{90},D_{92})$  & $0$               & $4140$ & $\mathbb{Z}_2^{45}$ \\
$(A_2,D_{81})$  & $\frac{81}{163}$ & $81 $ & $1$            & $(A_4,D_{67})$  & $\frac{134}{137}$ & $134 $ & $1$              & $(A_{92},D_{94})$  & $0$               & $4324$ & $\mathbb{Z}_2^{46}$ \\
$(A_2,D_{82})$  & $\frac{26}{55}$  & $82 $ & $\mathbb{Z}_2$ & $(A_4,D_{68})$  & $\frac{136}{139}$ & $136 $ & $1$              & $(A_{94},D_{96})$  & $0$               & $4512$ & $\mathbb{Z}_2^{47}$ \\
$(A_2,D_{83})$  & $\frac{83}{167}$ & $83 $ & $1$            & $(A_4,D_{69})$  & $\frac{46}{47}$   & $138 $ & $1$              & $(A_{96},D_{98})$  & $0$               & $4704$ & $\mathbb{Z}_2^{48}$ \\
$(A_2,D_{84})$  & $\frac{84}{169}$ & $84 $ & $1$            & $(A_4,D_{70})$  & $\frac{140}{143}$ & $140 $ & $1$              & $(A_{98},D_{100})$ & $0$               & $4900$ & $\mathbb{Z}_2^{49}$ \\
$(A_2,D_{85})$  & $\frac{9}{19}$   & $85 $ & $\mathbb{Z}_2$ & $(A_4,D_{71})$  & $\frac{26}{29}$   & $142 $ & $\mathbb{Z}_2^2$ &                    &                   &        &                     \\
$(A_2,D_{86})$  & $\frac{86}{173}$ & $86 $ & $1$            & $(A_4,D_{72})$  & $\frac{48}{49}$   & $144 $ & $1$              &                    &                   &        &                     \\
\end{longtable}
\end{center}

\begin{center}
 \renewcommand{\arraystretch}{1.25}
\begin{longtable}{c|c|c|c||c|c|c|c}
\caption{$(E_n,A_m)$ and $(E_n,D_m)$, with $n=6,7,8$ and $1\leq m\leq 100$. We called $h=24(c-a)$, $r$ the rank of the SCFT and $s$ the value of the $1$-form symmetry. We list only those theories with $0\leq h < 1$.}\\
$(E_n,A_m)$  & $h$ & $r$ & $s$ & $(E_n,D_m)$  & $h$ & $r$ & $s$ \\
    \hline
\endhead
\label{tab:GGpNHC}$(E_6,A_1)$        & $\frac{3}{7}$     & $3   $ &$1$              & $(E_6,D_3)$        & $0$               & $9   $ &  $\mathbb{Z}_3$   \\
                  $(E_6,A_3)$        & $0$               & $9   $ &$\mathbb{Z}_3$   & $(E_6,D_4)$        & $0$               & $12  $ &  $\mathbb{Z}_4$   \\
                  $(E_6,A_4)$        & $\frac{12}{17}$   & $12  $ &$1$              & $(E_8,D_3)$        & $\frac{12}{17}$   & $12  $ &  $1$              \\
                  $(E_6,A_6)$        & $\frac{18}{19}$   & $18  $ &$1$              & $(E_8,D_4)$        & $0$               & $16  $ &  $\mathbb{Z}_5$   \\
                  $(E_6,A_7)$        & $\frac{3}{5}$     & $21  $ &$\mathbb{Z}_3$   & $(E_8,D_6)$        & $0$               & $24  $ &  $\mathbb{Z}_3^2$ \\
                  $(E_7,A_2)$        & $\frac{5}{7}$     & $7   $ &$1$              &                    &                   &        &                   \\
                  $(E_7,A_4)$        & $\frac{14}{23}$   & $14  $ &$1$              &                    &                   &        &                   \\
                  $(E_7,A_6)$        & $\frac{21}{25}$   & $21  $ &$1$              &                    &                   &        &                   \\
                  $(E_7,A_8)$        & $0$               & $28  $ &$\mathbb{Z}_2^3$ &                    &                   &        &                   \\
                  $(E_8,A_1)$        & $\frac{1}{2}$     & $4   $ &$1$              &                    &                   &        &                   \\
                  $(E_8,A_2)$        & $\frac{8}{11}$    & $8   $ &$1$              &                    &                   &        &                   \\
                  $(E_8,A_3)$        & $\frac{12}{17}$   & $12  $ &$1$              &                    &                   &        &                   \\
                  $(E_8,A_5)$        & $0$               & $20  $ &$\mathbb{Z}_5$   &                    &                   &        &                   \\
                  $(E_8,A_6)$        & $\frac{24}{37}$   & $24  $ &$1$              &                    &                   &        &                   \\
                  $(E_8,A_9)$        & $0$               & $36  $ &$\mathbb{Z}_3^2$ &                    &                   &        &                   \\
                  $(E_8,A_{10})$     & $\frac{40}{41}$   & $40  $ &$1$              &                    &                   &        &                   \\
                  $(E_8,A_{11})$     & $\frac{4}{7}$     & $44  $ &$\mathbb{Z}_5$   &                    &                   &        &                   \\
                  $(E_8,A_{14})$     & $0$               & $56  $ &$\mathbb{Z}_2^4$ &                    &                   &        &                   \\
                  $(E_8,A_{19})$     & $\frac{4}{5}$     & $76  $ &$\mathbb{Z}_3^2$ &                    &                   &        &                   \\
\end{longtable}
\end{center}

\bibliographystyle{JHEP}
\bibliography{mybib}

\end{document}